\documentclass[preprint,10pt]{aastex}

\usepackage{multicol}
\usepackage{graphicx}
\usepackage{float}
\usepackage{enumerate}

\newcommand{\rr}{r}

\newcommand{\xiS}{\xi_S}

\newcommand{\exptime}{\tau_S}
\newcommand{\exptimeB}{\tau_B}
\newcommand{\nS}{n_S}
\newcommand{\nB}{n_B}
\newcommand{\lamS}{\lambda_S}
\newcommand{\lamB}{\lambda_B}
\newcommand{\rangeLamB}{\Lambda_B}

\newcommand{\thrtest}{{\cal{S}}}
\newcommand{\thresh}{\thrtest^\star}
\newcommand{\ulim}{{\cal{U}}}
\newcommand{\prob}{{\rm Pr}}
\newcommand{\rep}{^{\prime}}

\slugcomment{accepted for publication in ApJ}

\shorttitle{Upper Limits}
\shortauthors{Kashyap, van Dyk, Connors, Freeman, Siemiginowska, Xu, and Zezas}

\begin{document}

\title{On Computing Upper Limits to Source Intensities}

\author{
Vinay L.\ Kashyap\altaffilmark{1},
David A.\ van Dyk\altaffilmark{2},
Alanna Connors\altaffilmark{3},\\
Peter E. Freeman\altaffilmark{4},
Aneta Siemiginowska\altaffilmark{1},
Jin Xu\altaffilmark{2}, and
Andreas Zezas\altaffilmark{5}
}
\affil{$^1$Smithsonian Astrophysical Observatory, \\
60 Garden Street, Cambridge, MA 02138
\email{
vkashyap@cfa.harvard.edu \\
asiemiginowska@cfa.harvard.edu
}
}
\affil{$^2$ Department of Statistics, University of California, \\
Irvine, CA 92697-1250
\email{dvd@ics.uci.edu\\
jinx@ics.uci.edu
}
}
\affil{$^3$Eureka Scientific \\
{2452 Delmer Street Suite 100
 Oakland, CA 94602-3017}
\email{aconnors@eurekabayes.com}
}
\affil{$^4$ Department of Statistics, Carnegie Mellon University\\
5000 Forbes Avenue, Pittsburgh, PA 15213
\email{pfreeman@cmu.edu}
}
\affil{$^5$Physics Department, University of Crete, \\
P.O. Box 2208, GR-710 03, Heraklion, Crete, Greece
\email{azezas@cfa.harvard.edu}
}

\begin{abstract}

A common problem in astrophysics is determining how bright a source
could be and still not be detected in an observation. Despite the
simplicity with which the problem can be stated, the solution
involves complicated statistical issues that require careful analysis.
In contrast to the more familiar confidence bound, this concept has
never been formally analyzed, leading to a great variety of often
ad hoc solutions.  Here we formulate and describe the problem in a
self-consistent manner.  Detection significance is usually defined
by the acceptable proportion of false positives (background
fluctuations that are claimed as detections, or the Type~I error),
and we invoke the complementary concept of false negatives (real
sources that go undetected, or the Type~II error), based on the
statistical power of a test, to compute an upper limit to the
detectable source intensity.  To determine the minimum intensity
that a source must have for it to be detected, we first define a
detection threshold, and then compute the probabilities of detecting
sources of various intensities at the given threshold.  The intensity
that corresponds to the specified Type~II error probability defines
that minimum intensity, and is identified as the upper limit.  Thus,
an upper limit is a characteristic of the detection procedure rather
than the strength of any particular source.  It should not be
confused with confidence intervals or other estimates of a source
intensity. This is particularly important given the large number
of catalogs that are being generated from increasingly sensitive
surveys.  We discuss, with examples, the differences between these
upper limits and confidence bounds.  Both measures are useful
quantities that should be reported in order to extract the most
science from catalogs, though they answer different statistical
questions: an upper bound describes an inference range on the source
intensity, while an upper limit calibrates the detection process.
We provide a recipe for computing upper limits that applies to all
detection algorithms.

\end{abstract}


\section{Introduction}
\label{s:intro}

When a known or suspected source remains undetected at a prescribed
statistical significance during an observation, it is customary to
report the {\sl upper limit} on its intensity.  This limit is usually
taken to mean the largest intrinsic intensity that a nominal source
can have with an appreciable chance of going undetected.  Or
equivalently, it is the smallest intrinsic intensity it could have
before its detection probability falls below a certain threshold.
We emphasize that the upper limit is not meant to be an estimate
or even a bound on the intensity of the source, but rather it is a
quantification of the power of the {\it detection procedure} to
detect weak sources.  Thus, it is a measure that characterizes the
detection process.  The endpoints of the confidence interval, by
contrast, provide ranges of possible values for the source intensity
rather than quantify the sensitivity of the procedure.  While the
concept of upper limits has generally been well understood by
astronomers as a form of censoring (Isobe, Feigelson, \& Nelson
1986) with an intrinsic connection to the detectability of a source
(Avni et al.\ 1980), there has not been a statistically meaningful
description that encapsulates its reliance on detectability as well
as statistical significance.

Moreover, while the term ``upper limit'' has been traditionally
used in this manner, it has been also been used in cases where a
formal detection process is not applied (e.g., when a source is
known to exist at a given location because of a detection made in
some other wavelength).  In such cases, the upper edge of the
confidence interval is derived and noted as the upper limit,
regardless of the detectability of that source.  In order to prevent
confusion, we shall henceforth refer to the upper edge of a confidence
interval as the ``{\sl upper bound}.''  Despite the intrinsic
differences, numerous studies have described the computation of the
{\sl upper bound} as a proxy for the {\sl upper limit} in increasingly
sophisticated ways.  The parameter confidence interval is a
statistically well understood and is in common use.  Kraft et al.\
(1991) and Marshall (1992) for example applied Poisson Bayesian-likelihood
analysis to X-ray counts data to determine the credible range, and
thus set an upper bound, on the source intensity. Feldman \& Cousins
(1998) recognized that the classical confidence interval at a given
significance is not unique, and devised a scheme to determine
intervals by comparing the likelihoods of obtaining the observed
number of counts with the maximum likelihood estimate of the intensity
and a nominal intensity; this procedure produces unique intervals
where the lower edge overlaps with zero when there are very few
counts, and the upper edge stands as a proxy for an upper limit.
Variations in the background were incorporated via a sophisticated
Bayesian analysis by Weisskopf et al.\ (2007).

The similarity of nomenclature between upper limits and upper bounds
has led to considerable confusion in the literature on the nature
of upper limits, how to compute them, and what type of data to use
to do so.  Many techniques have been used to determine upper limits.
It is not feasible to list all of these,\footnote{ An ADS query on
astronomy abstracts, within the past year (excluding arXiv),
containing ``upper limit'', yields roughly two papers per day (759).
A quick survey shows this term used in several disparate ways: some
are upper bounds of confidence regions, often convolved with physics
information to get the upper bound of a confidence region on (say)
mass; others are clearly the theoretical power of a suggested test;
yet others use 'upper limits' from previous work to obtain, e.g.,
line slopes.} but for the sake of definiteness, we list a few methods
culled from the literature: the techniques range from using the
root-mean-square deviations in the background to set the upper limit
(Gilman et al.\ 1986, Ayres 1999, Perez-Torres, et al.\ 2009),
adopting the source detection threshold as the upper limit (Damiani
et al.\ 1997, Erdeve et al.\ 2009, Rudnick et al.\ 2009), computing
the flux required to change a fit statistic value by a significant
amount (Loewenstein et al.\ 2009), computing the $p$-value for the
significance of a putative detection in the presence of background
(Pease et al.\ 2006, Carrera et al.\ 2008), and identifying the
upper limit with the parameter confidence bound (Schlegel \& Petre
1993, Rolke et al., 2005, Hughes et al.\ 2007).  Here, we seek to
clarify these historically oft-used terms in a statistically rigorous
way.

Our goal here is to illustrate the difference between upper {\sl
limits} and upper {\sl bounds}, and to develop a self-consistent
description for the former that can be used with all extant detection
techniques.  Bounds and Limits describe answers to different
statistical questions, and usually both should be reported in
detection problems.  We seek to clarify their respective usage here.
We set out the requisite definitions and statistical foundations
in \S\ref{s:back}.  In \S\ref{s:ulim}, we discuss the critical role
played by the detection threshold in the definition of an upper
limit and compare upper limits with upper bounds of confidence
intervals.  In \S\ref{s:back} and \S\ref{s:ulim} we use a simple
Poisson detection problem as a running example to illustrate our
methods.  In \S\ref{s:snr}, we apply them to a signal-to-noise
detection problem.  Finally, we summarize in \S\ref{s:summary}.

\section{Statistical Background}
\label{s:back}

Here we begin by describing our notation, and then discuss the nuances of the familiar concepts of confidence intervals, hypothesis testing, and statistical power.  A glossary of the notation used is given in Table~\ref{tab:glossary}.

\begin{deluxetable}{ll}
\tablecolumns{2}
\tablecaption{Symbols and notation \label{tab:glossary}}
\tablehead{ \colhead{Symbol} & \colhead{Description} }
\startdata
$\nS$ & counts observed in source area \\ 
$\nB$ & counts observed in background area \\ 
$\lamS$ & source intensity \\ 
$\lamB$ & background intensity \\ 
$\rangeLamB$ & range in background intensity $\lamB$ \\ 
$\exptime$ & exposure time \\ 
$\exptimeB$ & exposure time for the background \\ 
$\rr$ & ratio of background to source area \\ 
$\thrtest$ & statistic for hypothesis test \\ 
$\thresh$ & detection threshold value of statistic $\thrtest$ \\ 
$\nS^\star$ & detection threshold value of statistic $n_S$ \\ 
$\ulim$ & upper limit \\ 
$\alpha$ & the maximum probability of false detection \\
$\beta$ & the probability of a detection \\
$\beta_{\rm min}$ & the minimum probability of detection of a source with $\lamS=\ulim$ \\ 
$\prob(.)$ & probability of \\ 
$n \sim f(.)$ & denoting that $n$ is sampled from the distribution $f(.)$ \\ 
${\rm Poisson}(\lambda)$ & Poisson distribution with intensity $\lambda$ \\
${\cal N}(\mu,\sigma)$ & Gaussian (i.e., normal) distribution with mean $\mu$ and variance $\sigma^2$ \\
\enddata
\end{deluxetable}

\subsection{Description of the Problem}

\label{s:desc}

Our study is carried out in the context of background-contaminated
detection of point sources in photon counting detectors, as in X-ray
astronomical data.  We set up the problem for the case of uncomplicated
source detection (i.e., ignoring source confusion, intrinsic
background variations, and instrumental effects such as vignetting,
detector efficiency, PSF structure, bad pixels, etc).  However, the
methodology we develop is sufficiently general to apply in complex
situations.

There is an important, subtle, and often overlooked distinction
between an upper limit and the upper bound of a confidence interval,
and the primary goal of this article is to illuminate this difference.
The confidence interval is the result of inference on the source
intensity, while the upper limit is a measure of the power of the
detection process.  We can precisely state this difference in the
context of an example.  Suppose that we have a typical case of a
source detection problem, where counts are collected in a region
containing a putative or possible source and are compared with
counts from a source-free region that defines the background.  If
the source counts exceed the threshold for detection, the source
is considered detected.  This detection threshold is usually
determined by limiting the probability of a false detection.  If
the threshold were lower, there would be more false detections.
Given this setup, we might ask how bright must a source be in order
to ensure detection.  Although statistically there are no guarantees,
the upper limit is the minimum brightness that ensures a certain
{\sl probability of detection}.  Critically, this value can be
computed before the source counts are observed.  It is based on two
probabilities, (1) the probability of a false detection which
determines the detection threshold, and (2) the minimum probability
that a bright source is detected.  Although the upper limit is
primarily of interest when the observed counts are less than the
detection threshold, it does not depend on the observed counts.
This is in sharp contrast to a confidence interval for the source
intensity, that is typically of the form {\sl ``source intensity
estimate plus or minus an error bar,''} where the estimate and the
error bars depend directly on the observed source counts in the
putative source region.  Of course, the functional form of the
confidence interval may be more complicated than in this example,
especially in low count settings, but any reasonable confidence
interval depends on the source counts, unlike upper limits.  The
fact that upper limits do not depend on the source counts while
confidence intervals do should not be viewed as an advantage of one
quantity or the other. Rather it reflects their differing goals.
Upper limits quantify the power of the detection procedure and
confidence intervals describe likely values of the source intensity.
These distinctions are highlighted with illustrative examples in
\S\ref{sec:illustex}.

To formalize discussion, suppose that a known source has an intrinsic
intensity in a given passband of $\lamS$ and that the background
intensity under the source is $\lamB$.  Further suppose that the
source is observed for a duration $\exptime$ and that $\nS$ counts
are collected, and similarly, a separate measurement of the background
could be made over a duration $\exptimeB$ and $\nB$ counts are
collected.  If the background counts are collected in an area $\rr$
times the source area,\footnote{
For clarity, we assume that the expected intensities are in units
of counts per unit time and that the source and background counts
are collected over pre-specified areas in an image.  However, our
analysis is not restricted to this scenario.  The nominal
background-to-source area ratio $\rr$ could include differences in
exposure duration and instrument effective area.  Furthermore, the
nominal exposure duration could also incorporate effective area,
e.g., to have units [photons~count$^{-1}$~cm$^2$~s], which implies
that the expected intensity $\lamS$ will have units
[photons~s$^{-1}$~cm$^{-2}$].  Regardless of the units of $\lamS$
and $\lamB$, the likelihood is determined by the Poisson distribution
on the expected and observed counts as in Equation~\ref{eq:basicpoi}.
}  we can relate the observed counts to the expected intensities,
\begin{eqnarray}
\nonumber
\nB | (\lamB,\rr,\exptimeB) &\sim& {\rm Poisson}(\rr \exptimeB \lamB) \\
\nS | (\lamS,\lamB,\exptime) &\sim& {\rm Poisson}\Big(\exptime(\lamS+\lamB)\Big) \,,
\label{eq:basicpoi}
\end{eqnarray}
in the background and source regions respectively, where $\nB$ and
$\nS$ are independent.  For simplicity, we begin by assuming that
$\lamB$ is known.

\subsection{Confidence and Credible Intervals}
\label{s:CI}

Confidence intervals give a set of values for the source intensity
that are consistent with the observed data.  They are typically
part of the inference problem for the source intensity.  The basic
strategy is to compute an interval of parameter values so that on
repeated observations a certain proportion of the intervals contain
the true value of the parameter.  It is in this ``repeated-observation''
sense that a classical confidence interval has a given probability
of containing the true parameter.  Bayesian credible intervals have
a more direct probabilistic interpretation.  They are computed by
deriving the posterior probability distribution of the source
intensity parameter given the observed counts and finding an interval
with nominal probability of containing the true rate  (see e.g.,
Loredo 1992, van Dyk et al.\ 2001, Kashyap et al.\ 2008). In summary,
confidence intervals are {\sl frequentist} in nature meaning that
they are interpreted in terms of repeated observations of the source.
Credible intervals, on the other hand, are {\sl Bayesian} in nature
meaning that they represent an interval of a certain posterior (or
other Bayesian) probability.

\subsubsection{Confidence Intervals}

From a frequentist point of view randomness stems only from data
collection---it is the data, not the parameters that are random.
Often we use a 95\% interval, but intervals may be at any {\sl
level} and we more generally refer to an $L$\% confidence interval.
Thus, the proper interpretation of a given interval is
\begin{quote}
{\sl $L$\% of experiments (i.e., observations) with intervals
computed in this way will result in intervals that contains the
true value of the source intensity.}
\end{quote}
In frequentist terms, this means that in any given experiment one
cannot know whether the true source intensity is contained in the
interval but if the experiment is repeated a large number of times,
{\sl about $L$\% of the resulting intervals will contain the true
value}.  Strictly speaking, the more colloquial understanding that
there is an $L$\% chance that the ``source intensity is contained
in the reported confidence interval,'' is incorrect.

Put another way, a confidence interval for the source intensity
gives values of $\lamS$ that are plausible given the observed counts.
Suppose that $\lamB=3$ and that for each value of $\lamS$, we
construct an interval ${\cal I}(\lamS)$ of possible values of the
source counts that has at least an $L$\% chance: $\prob(\nS \in
{\cal I}(\lamS)|\lamS,\lamB,\exptime) \geq L\%$.  Once the source
count is observed and assuming $\lamB$ is known, a confidence
interval can be constructed as the set of values of $\lamS$ for
which the observed count is contained in $\cal{I}(\lamS)$,
\begin{equation}
\{\lamS : \nS \in \cal{I}(\lamS)\} \,.
\end{equation}
In repeated observations, at least $L$\% of intervals computed in
this way cover the true value of $\lamS$.

\begin{figure*}[p]
\begin{center}
\includegraphics[width=6.5in]{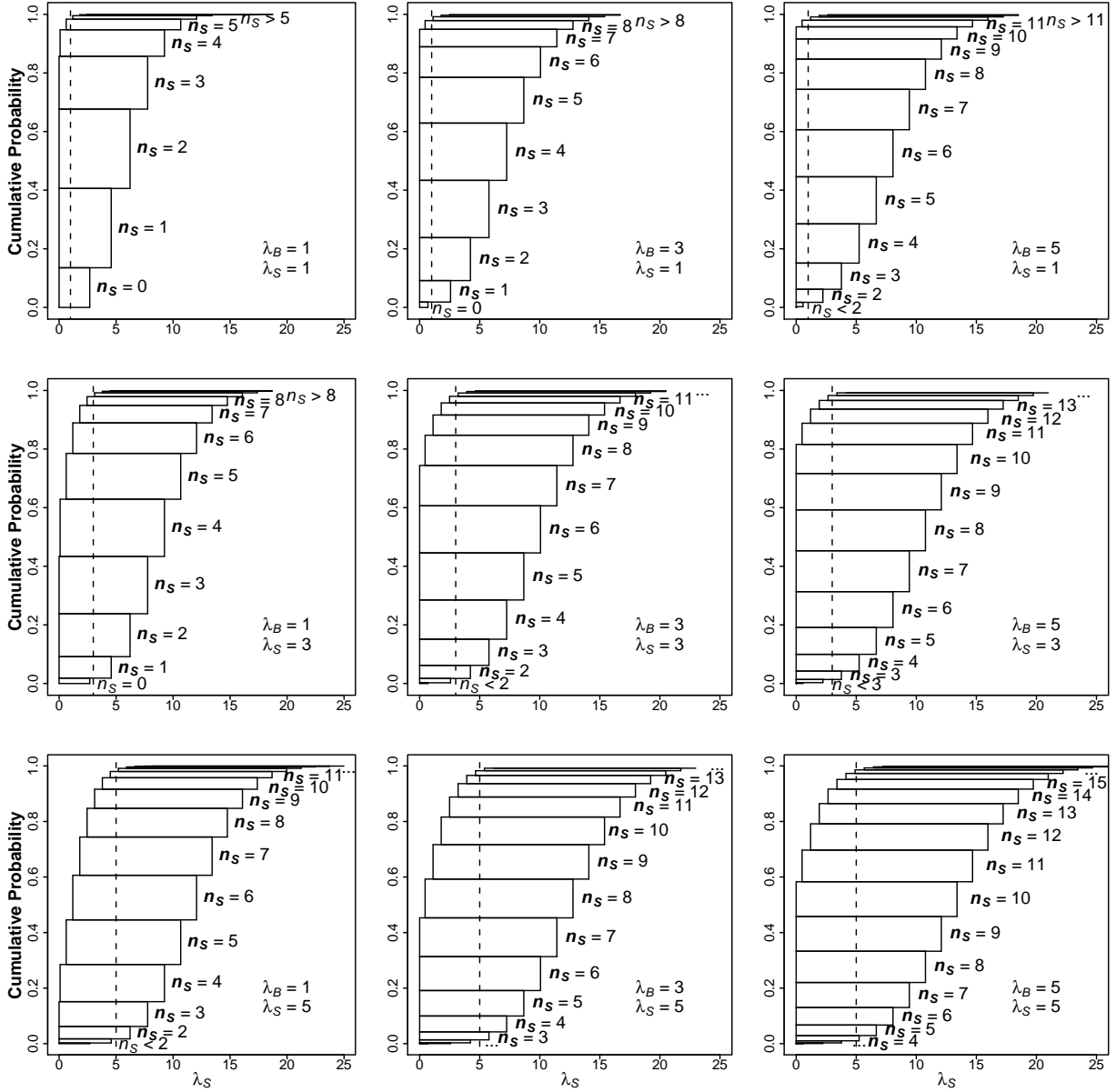} 
\caption{\small
Confidence Intervals for $\lamS$, Computed for Cases with Different
Source and Background Intensities.  The true value of $\lamS$ is
shown as a vertical dashed line and noted in the legend along with
the true value of $\lamB$.  We assume that $\lamB$ is known exactly
and adopt a nominal exposure $\exptime=1$ and background scaling
$\rr=10$.  Each box corresponds to a different value of $\nS$.  The
horizontal width of each box denotes the width of the 95\% confidence
interval, and the height denotes the probability of observing that
many counts given $\lamS$.
{\sl Top row:} $\lamS=1$ and for $\lamB=1,3,5$ for the left, middle, and right columns respectively.
{\sl Middle row:} As for the top row, for $\lamS=3$.
{\sl Bottom row:} As for the top row, for $\lamS=5$.
The figure illustrates that if the models are correctly specfied,
very short intervals should be rare.  Bayesian credible intervals
(not shown) look similar, at least in high count scenarios.}
\label{fig:CI}
\end{center}
\end{figure*}

The frequency coverage of confidence intervals is illustrated in
Figure~\ref{fig:CI}, where the confidence intervals of Garwood
(1936) for a Poisson mean are plotted as boxes of width equal to
the interval for various cases of observed counts (see van Dyk's
discussion in Mandelkern, 2002).  For given values of $\lamS$ and
$\lamB$ the probability of the possible values of the observed
counts can be computed using Equation~\ref{eq:basicpoi}; for each
of these possible values, there will be a different confidence
interval.  Thus, the confidence intervals themselves have their own
probabilities, which are represented as the heights of the boxes
in Figure~\ref{fig:CI}.  Because these are 95\% confidence intervals,
the cumulative heights of the boxes that contain the true value of
$\lamS$ in their horizontal range must be at least 0.95.  It is
common practice to only report confidence intervals for detected
sources so that only intervals corresponding to $\nS$ above some
threshold are reported.  Unfortunately this upsets the probability
that the interval contains  $\lamS$ upon repeated observations.
Standard confidence intervals are designed to contain the true value
of the parameter (say) 95\% of time, i.e., in 95\% of data sets.
If some of the confidence intervals are taken away (i.e., are not
reported because, e.g., the counts are too small), there is no
reason to expect that 95\% of those remaining will contain the true
value of the parameter. This is because instead of summing over all
values of $\nS$ to get a probability that exceeds 95\%, we are
summing only over those values greater than the detection threshold.
This results in a form of {\sl Eddington bias} and is discussed in
detail in \S\ref{s:eddington}.

\subsubsection{Credible Intervals}

In a Bayesian setting, probability is used to quantify uncertainty
in knowledge and in this regard parameters are typically viewed as
random quantities.  This distinction leads to a more intuitive
interpretation of the credible interval.  A credible interval at
the $L$\% level, for example, is any interval that contains the
true value of the parameter $L$\% of the time according to its
posterior distribution.  (See Park et al.~2008 for discussion on
interval selection.)  Thus, from a Bayesian perspective, it is
proper to say that there is an $L$\% chance that the source intensity
is contained in the reported credible interval. The corresponding
credible intervals look similar to the confidence intervals in
Figure~\ref{fig:CI}, at least in high count scenarios\footnote{In
small count scenarios Bayesian credible intervals may not exhibit
their nominal frequency coverage. They do however have the proper
Bayesian posterior probability.}.

So far we have considered a very simple problem with only one unknown
parameter, $\lamS$.  The situation is more complicated if there are
unknown {\sl nuisance parameters}, such as $\lamB$.  In this case,
frequency based intervals typically are constructed using asymptotic
arguments and/or by conditioning on ancillary statistics that yield
a conditional sampling distribution that does not depend on the
nuisance parameter.  Identifying ancillary statistics can be a
subtle task and the resulting intervals may not be unique.  Bayesian
intervals, can be constructed using a simple and clear  principle
known as {\sl marginalization}.  For example, if $\lamB$ is unknown,
the marginal posterior distribution of $\lamS$ is simply
\begin{eqnarray}
\nonumber p(\lamS | \nS, \nB,\exptime, \exptimeB, \rr) &=& \int p(\lamS, \lamB | \nS, \nB,\exptime,\exptimeB, \rr)~d\lamB \, \\
&=& \int p(\lamS | \lamB, \nS, \exptime) ~ p(\lamB | \nB, \exptimeB, \rr)~d\lamB \,.
\label{eq:marginlamB}
\end{eqnarray}
Credible intervals for $\lamS$ are computed just as before, but
using the marginal posterior distribution. \footnote{A popular
frequentist alternative to the marginal posterior distribution is
the profile likelihood function (see Park et al.\ 2008). Rather
than {\it averaging} over nuisance parameters, the profile likelihood
optimizes the likelihood over nuisance parameters for each value
of the parameter of interest.}

\subsection{Hypothesis Testing and Power}
\label{s:testing}

We emphasize that neither confidence nor credible intervals directly
quantify the detection sensitivity of an experiment.  To do this
we consider the detection problem in detail, which from a statistical
point of view is a test of the hypothesis that there is no source
emission in the given energy band\footnote{There is a close
relationship between confidence intervals and hypothesis testing.
If the interval includes zero, this indicates that there is a real
possibility of no source emission above the background and if the
source has not been otherwise detected there may be no source at
all.  Conversely, in Appendix~\ref{a:invert} we discuss how a
hypothesis test can be inverted to construct a confidence interval.},
i.e., a test of $\lamS=0$.  Formally, we test the {\sl null hypothesis}
that $\lamS=0$ against the {\sl alternative hypothesis} that
$\lamS>0$.  The test is conducted using a {\sl test statistic} that
we denote ${\thrtest}$.  An obvious choice for $\thrtest$ is the
counts in the source region, $\nS$; larger values of $\nS$ are
indicative of a detection of a source, since they become increasingly
less likely to have been obtained as a random fluctuation from the
background.  Other choices for $\thrtest$ are the signal-to-noise
ratio (as in the case of sliding-cell local-detect algorithms; see
\S\ref{s:snr}) or the value of the correlation of a counts image
with a basis function (as in the case of wavelet-based algorithms)
or a suitably calibrated likelihood-ratio test statistic (as in the
case of $\gamma$-ray detectors like {\sl Fermi}; see Mattox et al.\
1996).  The count in the source region is an example of a test
statistic that is stochastically increasing\footnote{
The term ``stochastically increasing'' means that there is a parameter
(here $\lamS$) that defines a distribution of observable values
(here $\nS$), and that all of the quantiles of $\nS$ increase as
$\lamS$ increases.  There is no guarantee that at any single instance
of observation, a higher $\lamS$ should lead to a higher $\nS$.
} in $\lamS$.  For any fixed $\thresh$, $\lamB$, $\exptime$,
$\exptimeB$ and $\rr$, the probability that the test statistic
$\thrtest$ is less than the threshold $\thresh$ decreases as $\lamS$
increases, i.e., $\prob({\thrtest} \leq \thresh |
\lamS,\lamB,\exptime,\rr)$ decreases as $\lamS$ increases.  We
assume that $\thrtest$ is stochastically increasing in $\lamS$
throughout.\footnote{
In principle, the test statistic is only required to have {\sl
different} distributions under the alternative and null hypotheses.
For simplicity, we assume it tends to be larger under the alternative.
}

\begin{figure}[t]
\begin{center}
{\includegraphics[width=2.9in]{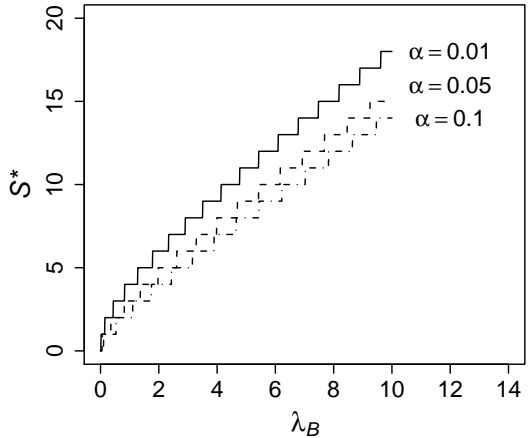}} 
\caption{
$\alpha$-Level Detection Threshold $\thresh$ as a Function of the
Background Intensity $\lamB$, for the Given $\alpha$ Levels.  Note
that this is calculated assuming that the source intensity $\lamS=0$.
The detection threshold increases with increasing $\lamB$ for a
given $\alpha$, and increases with decreasing $\alpha$ for a given
$\lamB$.
}
\label{fig:thresh}
\end{center}
\end{figure}

Because larger values of $\thrtest$ indicate a source, we need to
determine how large $\thrtest$ must be before we can declare a
source detection.  This is the done by limiting the probability of
a false detection, also known as a Type~I error.  Thus, the detection
threshold $\thresh$ is the smallest value such that
\begin{equation}
\prob(\thrtest > \thresh | \lamS=0,\lamB,\exptime,\exptimeB,\rr)\leq \alpha,
\label{eq:alpha}
\end{equation}
where $\alpha$ is the maximum allowed probability of a false
detection\footnote{
\label{ft:poialpha}
In the simple Poisson counts case, the probability of a Type~I error
given in Equation~\ref{eq:alpha} is computed as,
\begin{displaymath}
\prob(\thrtest > \thresh | \lamS=0,\lamB,\exptime,\exptimeB,\rr) =
\sum
\left\{
{e^{-\exptime\lamB} (\exptime\lamB)^{\nS\rep} \over \nS\rep!} \cdot
{e^{-\rr\exptimeB\lamB} (\rr\exptimeB\lamB)^{\nB\rep} \over \nB\rep!}
\right\},
\end{displaymath}
where the summation is over the set of values of $(\nS\rep,\nB\rep)$
such that $\thrtest(\nS\rep,\nB\rep) >\thresh$ and we substitute
$\lamS=0$ into the mean of $\nS$ given in Equation~\ref{eq:basicpoi}.
Each term in the summation is a product of the likelihood of obtaining
the specified counts in the absence of a source, given the background
intensity and other observational parameters.
 In the simple case where $\thrtest$ is the counts in the source
 region, $\nS$, and $\lamB$ is known (i.e., $\nB$ is not measured),
 this reduces to
\begin{eqnarray}
\nonumber\prob(\thrtest > \thresh | \lamS=0,\lamB,\exptime) 
&=& \sum_{\nS\rep=\thresh+1}^{\infty} \frac{e^{-\exptime\lamB}~(\exptime\lamB)^{\nS\rep}}{\Gamma(\nS\rep+1)} \\
\nonumber &=& 1 - \sum_{\nS\rep=0}^{\thresh} \frac{e^{-\exptime\lamB}~(\exptime\lamB)^{\nS\rep}}{\Gamma(\nS\rep+1)} \\ 
\nonumber &=& \frac{\gamma(\thresh+1,\exptime\lamB)}{\Gamma(\thresh+1)},
\end{eqnarray}
where
\begin{eqnarray}
\nonumber \gamma(\thresh+1,\exptime\lamB) &=& \Gamma(\thresh+1)~\int_{0}^{\exptime\lamB} e^{-t}~t^{\nS}~dt
\end{eqnarray}
is the incomplete gamma function (see Equations 8.350.1 and 8.352.1
of Gradshteyn \& Ryzhik (1980)).  In large count scenarios we may
use continuous Gaussian distributions with their variances equal
to their means in place of the discrete Poisson distributions in
Equation~\ref{eq:basicpoi}, see Equation~\ref{eq:snr-model} in
\S\ref{s:snr}.  In this case, we compute
\begin{displaymath}
\nonumber\prob(\thrtest > \thresh | \lamS=0,\lamB,\exptime,\exptimeB,\rr) =
\int
\left\{
{\exp\left[-{(\nS\rep -\exptime\lamB)^2\over \exptime\lamB}\right] \over \sqrt{2\pi\exptime\lamB}} \cdot
{\exp\left[-{(\nB\rep -\rr\exptimeB\lamB)^2\over \rr\exptimeB\lamB}\right] \over \sqrt{2\pi\rr\exptimeB\lamB}}
\right\}
d\nS\rep d\nB\rep,
\end{displaymath}
where the integral is over the region of values of $(\nS\rep,\nB\rep)$
such that $\thrtest(\nS\rep,\nB\rep) >\thresh$.
} and we declare a detection if the observed value of $\thrtest$
is {\it strictly greater than} $\thresh$:
\begin{quote}\sl
If $\thrtest \leq \thresh$ we conclude there is insufficient evidence
to declare a source detection.  If $\thrtest > \thresh$ we conclude
there is sufficient evidence to declare a source detection.
\end{quote}
We call $\thresh$ the {\sl $\alpha$-level detection threshold} and
sometimes write $\thresh(\alpha)$ to emphasize its dependance on
$\alpha$ (see Figure~\ref{fig:thresh}).
Note that $\alpha$ is a bound on the probability of a Type~I error;
the actual probability of a Type~I error is given by the probability
on the left-hand side of Equation~\ref{eq:alpha}. Due to the discrete
nature of the Poisson distribution, the bound is generally not
achieved and the actual probability of a Type-I error is less than
$\alpha$.

\begin{figure*}[t]
\begin{center}
\includegraphics[width=6.5in]{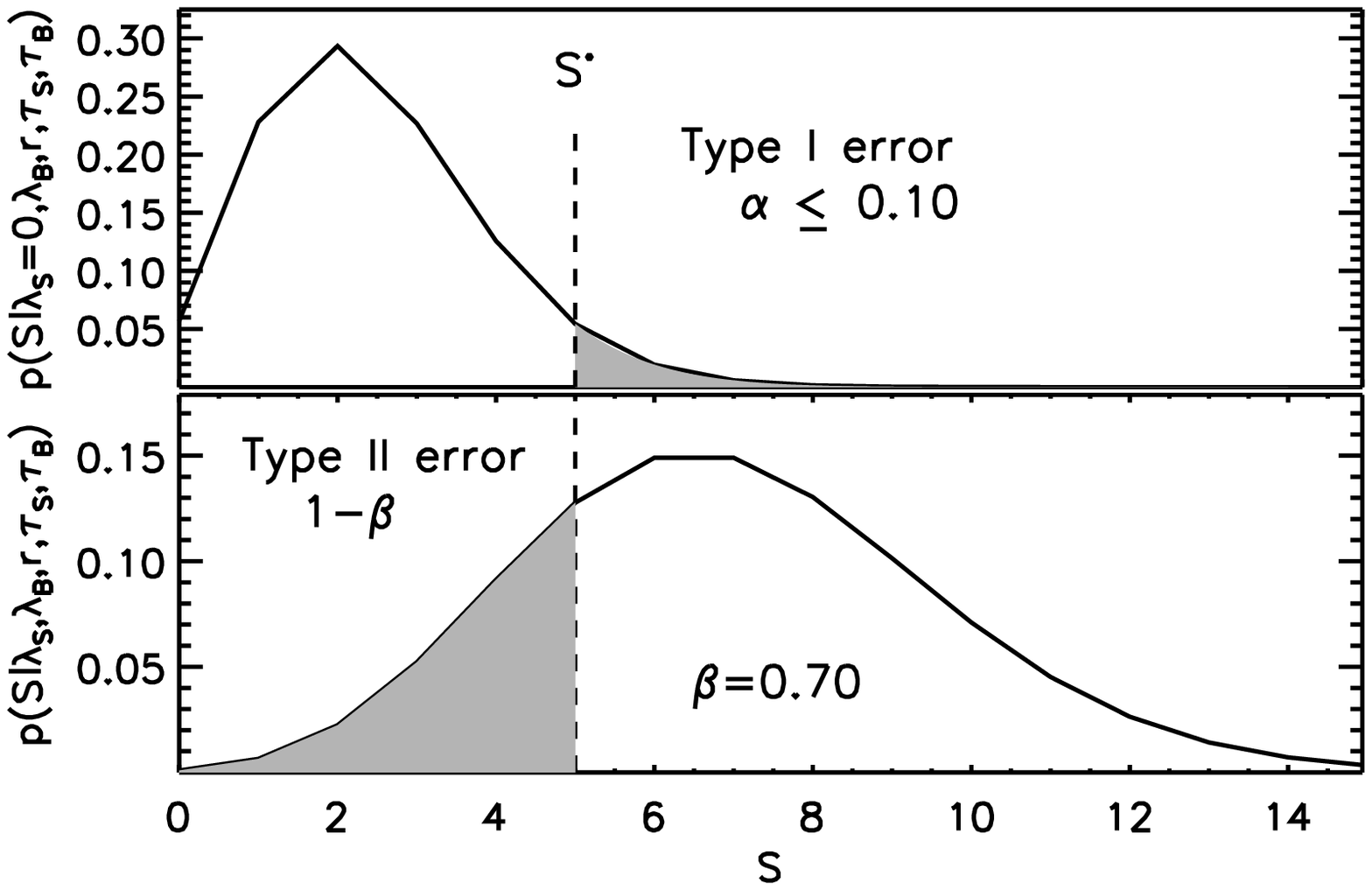} 
\caption{
Illustration of Type I and Type II errors.  A sketch of the probability
distribution of test statistic $\thrtest$ for specified values of
the source and background intensities is shown for the simple case
where $\thrtest\equiv\nS$ and the background is known (see
footnotes~\ref{ft:poialpha},\ref{ft:poibeta}).
The top panel depicts the probability
$\prob(\thrtest=\nS|\lamS=0,\lamB=2,\exptime=1)$ and the bottom
panel shows $\prob(\thrtest=\nS|\lamS=5,\lamB=2,\exptime=1)$.  The
vertical dashed line is a nominal detection threshold $\thresh$
that corresponds to a significance of $\alpha \le 0.1$, i.e.,
$\thresh=5$.  The Type~I error, or the probability of a false
positive, is shown by the shaded region to the right of the threshold.
(The actual Type-I error for the adopted parameters is $0.05$;
values of $\thresh$ less than 5 will cause the Type-I error to
exceed the specified significance.)  The Type~II error is the
probability of a false negative and is shown (for $\lamS=5$) by the
shaded region to the left of the threshold in the lower panel.  The
detection probability of a source with intensity $\lamS=5$ is
$\beta=0.7$ for this choice of $\alpha$, for the given background
intensity, and for the exposure time.
}
\label{fig:alfabetillus}
\end{center}
\end{figure*}

Although its role in the definition of the detection threshold
indicates that it is viewed as the more important concern, a {\sl
false detection}, also known as a ``false positive,'' is not the
only type of error.  A {\sl false negative}, or Type~II error,
occurs when a real source goes undetected (see
Figure~\ref{fig:alfabetillus}).  The probability of a false negative
is quantified through the {\sl power} of the test to detect a source
as a function of its intensity,
\begin{equation}
\beta(\lamS) = \prob(\thrtest > \thresh | \lamS,\lamB, \exptime, \exptimeB, \rr) \,.
\label{eq:beta}
\end{equation}
Equation~\ref{eq:beta} gives the probability of a detection.  For
any $\lamS>0$, this is the power of the test or one minus the
probability of a false negative.\footnote{\label{ft:poibeta}
Here we use $\beta$ to represent the power of the test, or one minus
the probability of a Type II error.  The statistical literature
uses the notation $\beta$ to denote either the Type II error (i.e.,
accepting the null hypothesis when it is false; e.g., Eadie et al.\
1971) or for the power of the test itself (as we have done here;
e.g., Casella \& Berger 2002).  As in the case of calculating
$\alpha$ (see footnote~\ref{ft:poialpha}), we can calculate $\beta$
as
\begin{displaymath}
\prob(\thrtest > \thresh | \lamS, \lamB, \exptime, \exptimeB, \rr) =
\sum
\left\{
{e^{-\exptime(\lamS+\lamB)} (\exptime (\lamS+\lamB))^{\nS\rep} \over \nS\rep!} \cdot
{e^{-\rr\exptimeB\lamB} (\rr\exptimeB\lamB)^{\nB\rep} \over \nB\rep!}
\right\},
\end{displaymath}
where again the summation is over the set of values of $(\nS\rep,\nB\rep)$
such that $\thrtest(\nS\rep,\nB\rep)>\thresh$.  In the simple case
where $\thrtest=\nS$, $\lamB$ is known, and $\nB$ is not measured,
we find
\begin{displaymath}
\prob(\thrtest>\thresh|\lamS,\lamB,\exptime) = \frac{\gamma(\thresh+1,\exptime (\lamS+\lamB))}{\Gamma(\nS+1)} \,.
\end{displaymath}
}   For $\lamS=0$, Equation~\ref{eq:beta} gives the probability of
a false detection (cf.\ Equation~\ref{eq:alpha}) and consequently,
$\beta(0) \leq\alpha$.  This reflects the trade-off in any detection
algorithm: the compromise between minimizing the number of false
detections against maximizing the number of true detections.  That
is, if the detection threshold is set low enough to detect weaker
sources, the algorithm will also produce a larger number of false
positives that are actually background fluctuations.  Conversely,
the more stringent the criterion for detection, the smaller the
probability of detecting a real source (this is illustrated by the
location of the threshold $\thresh$ that defines both $\alpha$ and
$\beta$ in Figure~\ref{fig:alfabetillus}).  Note that although our
notation emphasizes the dependence of the power on $\lamS$, it also
depends on $\lamB$, $\exptime$, $\exptimeB$, and $\rr$.

\begin{figure*}[t]
\begin{center}
\includegraphics[width=6.5in]{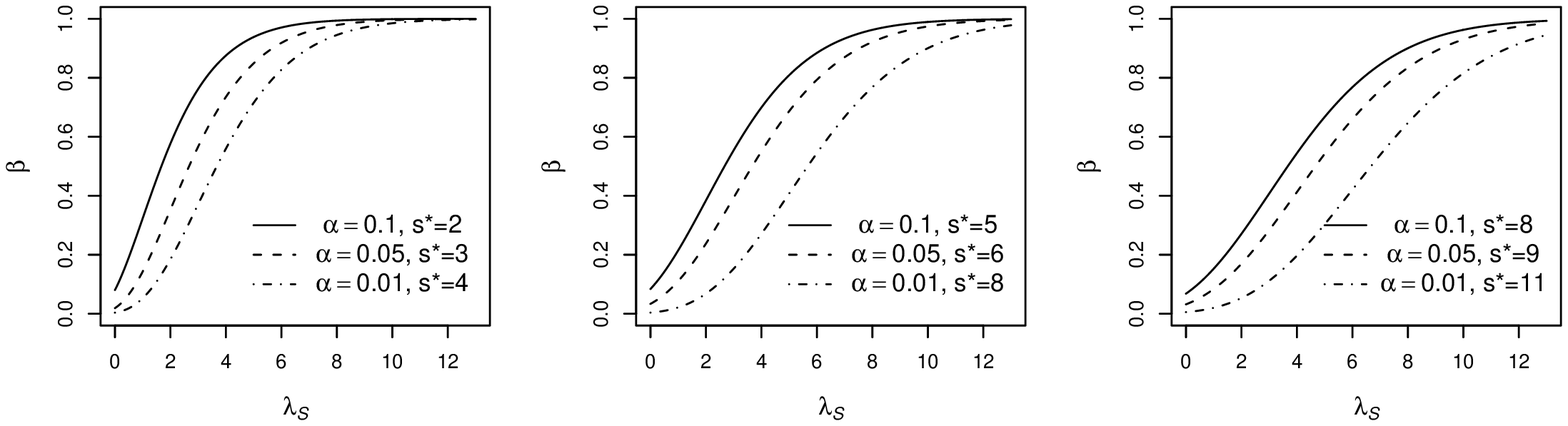} 
\caption{
Power of the Test, $\beta$, to Detect a Source as a Function of
Source Intensity, $\lamS$, and Detection Threshold, $\thresh$.  The
curves are calculated for different values of the background intensity
(the same values as in Figure~\ref{fig:CI}), $\lamB=1$ (left),
$\lamB=3$ (middle), and $\lamB=5$ (right).  The individual curves
show $\beta(\lamS)$ for different $\thresh$, each of which corresponds
to a different bound on the probability of a Type~I error,  $\alpha$,
see Figure~\ref{fig:thresh}.  The solid, dashed, and dash-dotted
lines correspond to increasing detection thresholds,  and decreasing
values of $\alpha$.  As one would expect, $\beta$ is higher for
larger $\lamS$ and lower $\lamB$, i.e., if the source is stronger
or the background is weaker, it is easier to detect.
}
\label{fig:power}
\end{center}
\end{figure*}

The power calculation is shown for the simple Poisson case in
Figure~\ref{fig:power}, where $\beta(\lamS)$ is plotted for different
instances of $\lamB$ and for different levels of the detection
threshold $\thresh$.  As expected, stronger sources are invariably
detected.  For a given source intensity, an increase in the background
or a larger detection threshold (i.e., lower $\alpha$) both cause
the detection probability to decrease.  In a typical observation,
the background and the detection threshold are already known, and
thus it is possible to state precisely the intensity $\lamS$ at
which the source will be detected at a certain probability.  We may
set a certain minimum probability, $\beta_{\rm min}$ of detecting
a ``bright'' source by setting the exposure time long enough so
that any source with intensity greater than a certain pre-specified
cutoff has probability $\beta_{\rm min}$ or more of being detected.
Conversely, we can determine how bright a source must be in order
to have probability $\beta_{\rm min}$ or more of being detected
with a given exposure time.  This allows us to define an upper limit
on the source intensity by setting a minimum probability of detecting
the source.  This latter calculation is the topic of \S~\ref{s:def}
and the basis of our definition of an upper limit.

Power calculations are generally used to determine the minimum
exposure time required to ensure a minimum probability of source
detection (see Appendix~\ref{a:ul-pow}). In \S\ref{s:ulim} we use
them to construct upper limits.

\section{Upper Limits}
\label{s:ulim}

In this section, we develop a clear statistical definition of an
upper limit  that (i) is based on well-defined principles, (ii)
depends only on the method of detection, (iii) does not depend on
prior or outside knowledge about the source intensity, (iv) corresponds
to precise probability statements, and  (v) is internally self-consistent
in that all values of the intensity below the upper limit are less
likely to be detected at the specified Type-I error rate and values
above are more likely to be detected.

\subsection{Definition}
\label{s:def}

In astronomy upper limits are inextricably bound to source detection:
by an upper limit, an astronomer means
\begin{quote}{\sl
The maximum intensity that a source can have without having at least
a probability of $\beta_{\rm min}$ of being detected under an
$\alpha$-level detection threshold. }
\end{quote}
or conversely,
\begin{quote}{\sl
The smallest intensity that a source can have with at least a
probability of $\beta_{\rm min}$ of being detected under an
$\alpha$-level detection threshold.}
\end{quote}
Unlike a confidence interval, the upper limit depends directly on
the detection process and in particular on the maximum probability
of a false detection and the minimum power of the test, that is on
$\alpha$ and $\beta_{\rm min}$ respectively.  In this way, an upper
limit incorporates both the probabilities of a Type I and a Type
II error.  Formally, we define the upper limit, ${\ulim}(\alpha,
\beta_{\rm min})$ to be the smallest $\lamS$ such that
\begin{equation}
\prob( {\thrtest} > \thresh(\alpha) | \lamS,\lamB, \exptime, \exptimeB, \rr)\geq\beta_{\rm min}.
\label{eq:power}
\end{equation}
Commonly used values for $\beta_{\rm min}$ throughout statistics are
0.8 and 0.9.
If $\beta_{\rm min} \approx1$, $\ulim(\alpha, \beta_{\rm min})$
represents the intensity of a source that is unlikely to go undetected,
and we can conclude that an undetected source is unlikely to have
intensity greater than $\ulim(\alpha, \beta_{\rm min})$.

\begin{figure}[t]
\begin{center}
\includegraphics[width=2.9in]{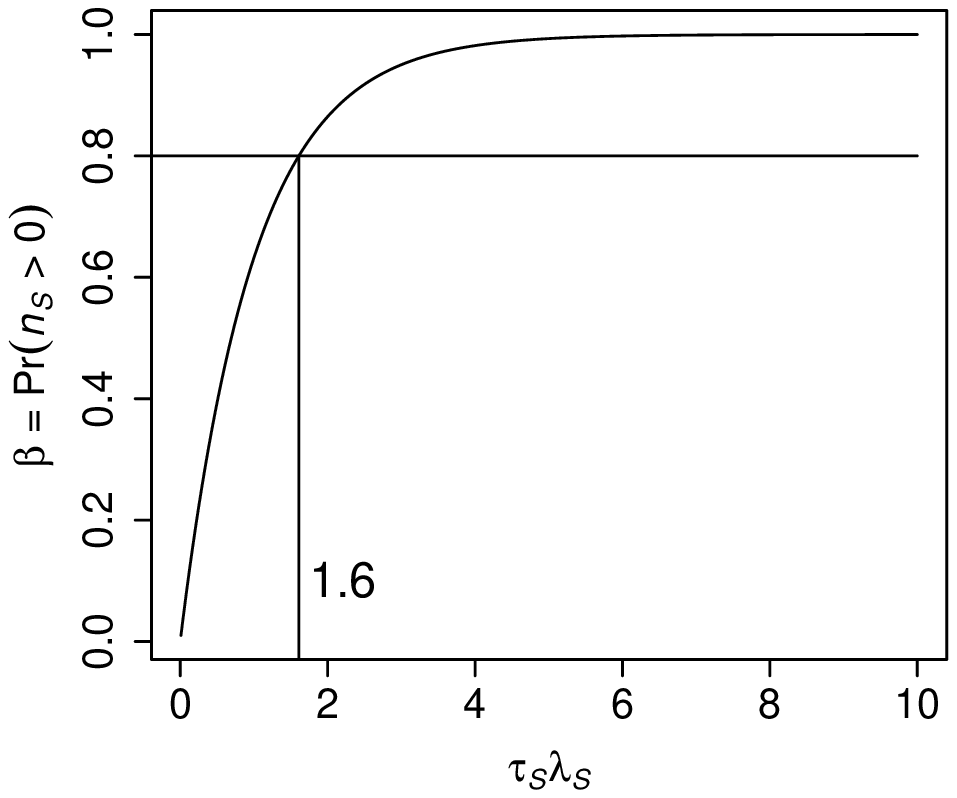} 
\caption{
Upper Limit with No Background Contamination.
The figure plots $\prob(\nS > 0 | \lamS, \lamB=0, \exptime)$ as a
function of $\exptime\lamS$, thus giving $\ulim(\alpha, \beta_{\rm
min})$ for any given $\exptime$ and every value of $\beta_{\rm
min}$.
For example, reading across the line plotted at $\beta=0.8$,  gives
$\exptime \ulim(\alpha=0.05, \beta_{\rm min}=0.8)= 1.6$, which can
be solved for the upper limit for any value of $\exptime$.
Notice the upper limit decreases in inverse proportion to $\exptime$.
}
\label{fig:noback}
\end{center}
\end{figure}

The simplest example occurs in the hypothetical situation when
$\lamB$ is known to be zero  and there is no background observation.
In this case we set $\thrtest=\nS$ and note that $\prob(\nS > 0 |
\lamS=0,\lamB=0,\exptime) =0$ so the detection threshold is zero
counts and we declare a detection if there is even a single count.
(Recall, we declare a detection only if $\thrtest$ is strictly
greater than $\thresh$.)  The upper limit in this case is the
smallest value of $\lamS$ with probability  of detection greater
than $\beta_{\rm min}$.
Figure~\ref{fig:noback} plots $\prob(\nS > 0 | \lamS, \lamB=0,
\exptime)$ as a function of $\exptime\lamS$, thus giving $\ulim(\alpha,
\beta_{\rm min})$ for any given $\exptime$ and every value of
$\beta_{\rm min}$.  Notice the upper limit decreases in inverse
proportion to $\exptime$.

When $\lamB$ is greater than zero but  well-determined and can be
considered to be known, the detection threshold using $\thrtest=\nS$
is given in Equation~\ref{eq:alpha}.  With this threshold in hand
we can determine the maximum intensity a source can have with
significant probability of not producing a large enough fluctuation
above the background for detection.  This is the upper limit.

\begin{figure*}[p]
\begin{center}
\includegraphics[width=6.5in]{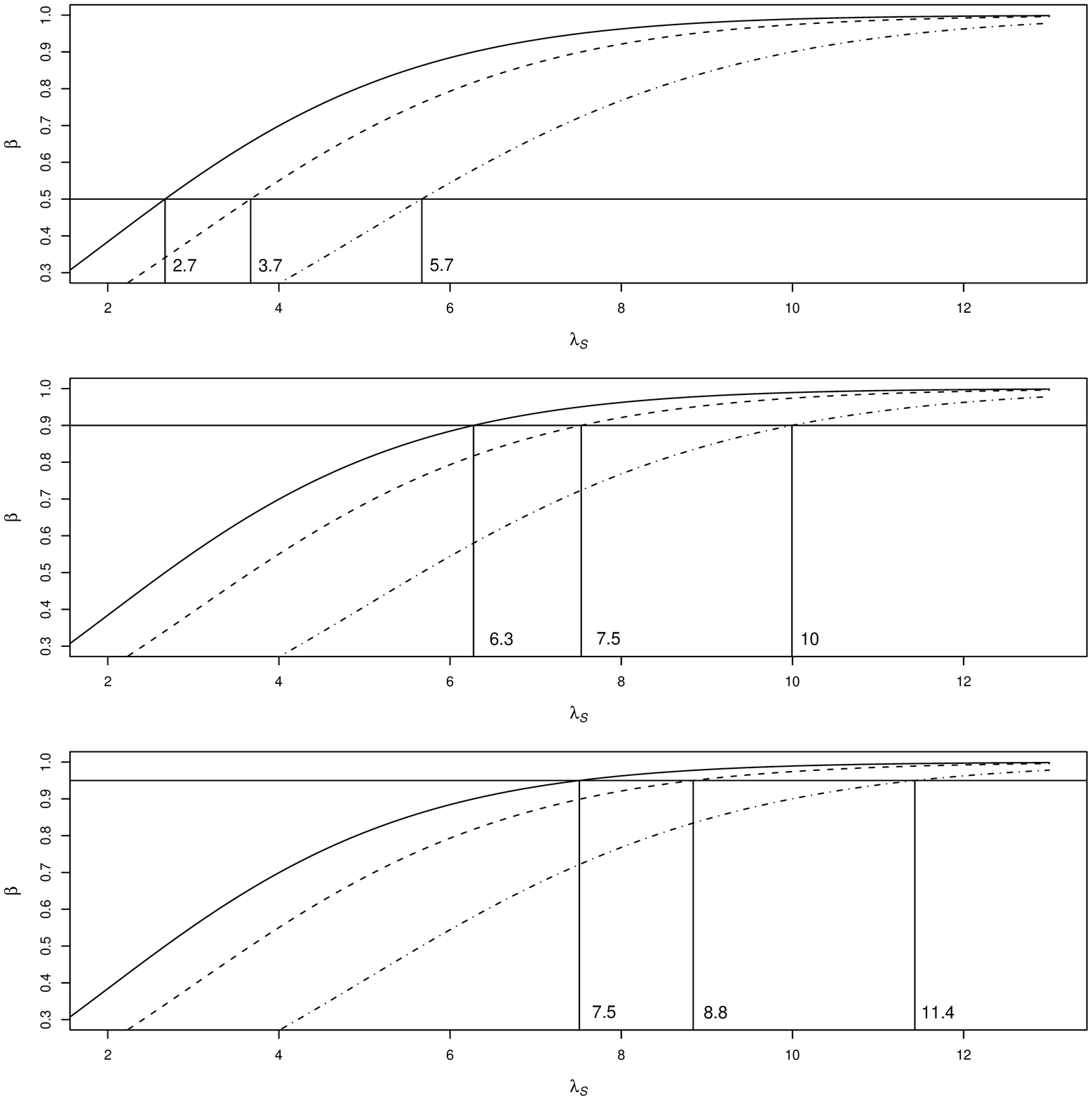} 
\caption{
Computing Upper Limits Based on the Probability of Detecting a
Source.  The figure illustrates how upper limits may be defined for
different probabilities of source detection under a given detection
threshold.  The curves correspond to $\beta(\lamS)$ for different
values of $\thresh$ and $\alpha$: 5 and 0.1 (solid), 6 and 0.05
(dashed), and 8 and 0.01 (dash-dotted) and were all computed with
$\lamB=3$, as in the middle panel of Figure~\ref{fig:power}.  Upper
limits are computed by first adopting an acceptable probability for
a source detection, and then computing the intercept on $\lamS$ of
the $\beta(\lamS)$ curves.  The panels show the value of the upper
limits for the different values of  $\thresh$ for $\beta_{\rm
min}=0.5$ (top), $\beta_{\rm min}=0.9$ (middle), and $\beta_{\rm
min}=0.95$ (bottom).
}
\label{fig:ulim}
\end{center}
\end{figure*}

In particular, $\ulim(\alpha, \beta_{\rm min})$ is the largest value
of $\lamS$ such that $\prob( \nS \leq \thresh(\alpha) | \lamS,\lamB,
\exptime) > 1-\beta_{\rm min}$.  This is illustrated for three
different values of $\beta_{\rm min}$ (panels) and three different
values of $\alpha$ (line types) in Figure~\ref{fig:ulim}.  Notice
that the upper limit increases as $\beta_{\rm min}$ increases and
as $\alpha$ decreases.

\subsubsection{Illustrative Examples}
\label{sec:illustex}

To illustrate the difference between confidence bounds and upper
limits, in the context of a detection process, we consider two
simple examples.  The first is an extreme case where the background
intensity is known to be identically zero, and even one count in
the source region would be classified as a detection.  In this case
the upper limit is the smallest source intensity that can produce
one count at a specified probability, e.g., a source with intensity
of 5 generates one or more counts at a probability of $\approx99.7$\%
(see \S\ref{s:def}).  In contrast, if one count is seen in the
source region, the upper bound of an equal-tail 99.7\% interval on
the source intensity is $8.9$ (Gehrels 1986).  Thus, while similar
in magnitude, it can be seen that upper bounds and upper limits are
different quantities, describing different concepts.

Second, consider a more realistic case where the background is
measured in a large region thought to be free of sources and scaled
to the area covered by the source.  Suppose that 800 counts are
observed in an area 400 times larger than the source area, and 3
counts are seen in the putative source region itself.  The credible
interval for the source intensity may be calculated at various
significance levels (\S\ref{s:CI}; see also van Dyk et al.\ 2001),
and for this case we find that the 68\% credible interval with the
lower bound at $0$ is $[0,2.1]$, and the 99.7\% interval is $[0,8.3]$.
But the question then arises as to whether the counts seen in the
source region are consistent with a fluctuation of the observed
background or not.  Since at a minimum 7 counts are needed for a
detection at a probability of 0.997 (corresponding to a Gaussian-equivalent
``$3\sigma$'' detection), it is considered that the source is not
detected.  The question then becomes how bright the source would
have to be in order to be detected with a certain probability.
Since a source of intensity 5.7 would have a 50\% probability of
producing sufficient counts for a detection, this sets an {\sl upper
limit} $\ulim(\alpha=0.003,\beta_{\rm min}=0.5)=5.7$ counts on the
undetected source's intensity (for a Type I error $\alpha=0.003$
and a Type II error $\beta=0.5$; see \S\ref{s:def}).  Note that
this limit is the same regardless of how many counts are actually
seen within the source region, as expected from a quantity that
calibrates the detection process.  In contrast, the inference on
the source intensity is always dependent on the number of observed
source counts.

\subsection{Unknown Background Intensity}
\label{s:unknownback}

So far our definition of an upper limit assumes that there are no
unknown nuisance parameters, and in particular that $\lamB$ is
known.  Unfortunately, the probabilities in Equations~\ref{eq:alpha}
and \ref{eq:power} cannot be computed if $\lamB$ is unknown.  In
this section we describe several strategies that can be used in the
more realistic situation when $\lamB$ is not known precisely.

The most conservative procedure ensures that the detection probability
of the upper limit is greater than $\beta_{\rm min}$ for any possible
value of $\lamB$.  Generally speaking, the larger $\lamB$ is, the
larger $\lamS$ must be in order to be detected with a given
probability, and thus the larger the upper limit.  Thus, a useful
upper limit requires a finite range, $\rangeLamB$, to be specified
for $\lamB$.  Given this range, a {\sl conservative upper limit}
can be defined as the smallest $\lamS$ that satisfies
\begin{equation}
\inf_{\lamB \in \rangeLamB} \prob( {\thrtest} > \thresh(\alpha) | \lamS,\lamB, \exptime, \exptimeB, \rr) \ \geq \ \beta_{\rm min}.
\label{eq:ULinf}
\end{equation}
(We use the term {\sl infimum} ($\inf$) rather than {\sl minimum}
to allow for the case when the minimum may be on the boundary of,
but outside, the range of interest.  It is the largest number that
is smaller than all the numbers in the range.  For instance, the
minimum of the range $\{x>0\}$ is undefined, but the infimum is
$0$.)  Unfortunately, unless the range of values $\rangeLamB$ is
relatively precise, this upper limit will often be too large to be
useful.

In practice, there is better solution.  The background count provides
information on the likely values of $\lamB$ that should be used
when computing the upper limit.  In particular the distribution of
$\lamB$ given $\nB$ can be computed using standard Bayesian
procedures\footnote{In the presence of a nuisance parameter,
frequentist procedures are more involved and typically require
conditioning on an ancillary statistic, see, e.g., Appendix~\ref{a:bkgd}.
In the Bayesian case, the posterior distribution,
$p(\lamB|\nB,\exptimeB,\rr) \propto p(\lamB)~p(\nB|\lamB,\exptimeB,\rr)$,
is the product of a prior distribution and the likelihood, normalized
so that the posterior distribution integrates to 1.  There are many
choices of prior distributions available for $\lamB$, ranging from
uniform on $\lamB$, to $\gamma$, to uniform in $\log(\lamB)$ (see,
e.g., van Dyk et al.\ 2001).} and used to evaluate the {\sl expected
detection probability},
\begin{equation}
\beta(\lamS) = \int \prob( {\thrtest} > \thresh(\alpha) | \lamS,\lamB, \exptime,\exptimeB, \rr) p(\lamB | \nB, \exptimeB,\rr)d\lamB,\label{eq:ULpost}
\end{equation}
where $\thresh(\alpha)$ is the smallest value such that
\begin{equation}
\int\prob(\thrtest  > \thresh(\alpha) | \lamS=0,\lamB,\exptime,\exptimeB, \rr) p(\lamB | \nB, \exptimeB, \rr)d\lamB \ \leq \ \alpha.
\label{eq:alpha2}
\end{equation}
The upper limit is then computed as the smallest $\lamS$ that
satisfies $\beta(\lamS)  \geq  \beta_{\rm min}$.  Unlike the upper
limit described in \S\ref{s:def}, these calculations require data,
in particular $\nB$.  For this reason, we call the smallest $\lamS$
that satisfies Equation~\ref{eq:ULpost} the {\sl background count
conditional upper limit} or {\sl bcc upper limit}.  An intermediate
approach that is more practical than using Equation~\ref{eq:ULinf}
but more conservative than using Equation~\ref{eq:ULpost} is to
simply compute a high percentile of $p(\lamB | \nB)$, perhaps  its
95th percentile.  The procedure for known $\lamB$ can then be used
with this percentile treated as the known value of $\lamB$.  This
is a conservative strategy in that it assumes a nearly worst case
scenario for the level of background contamination.

\begin{figure*}[t]
\begin{center}
\includegraphics[width=4.3in]{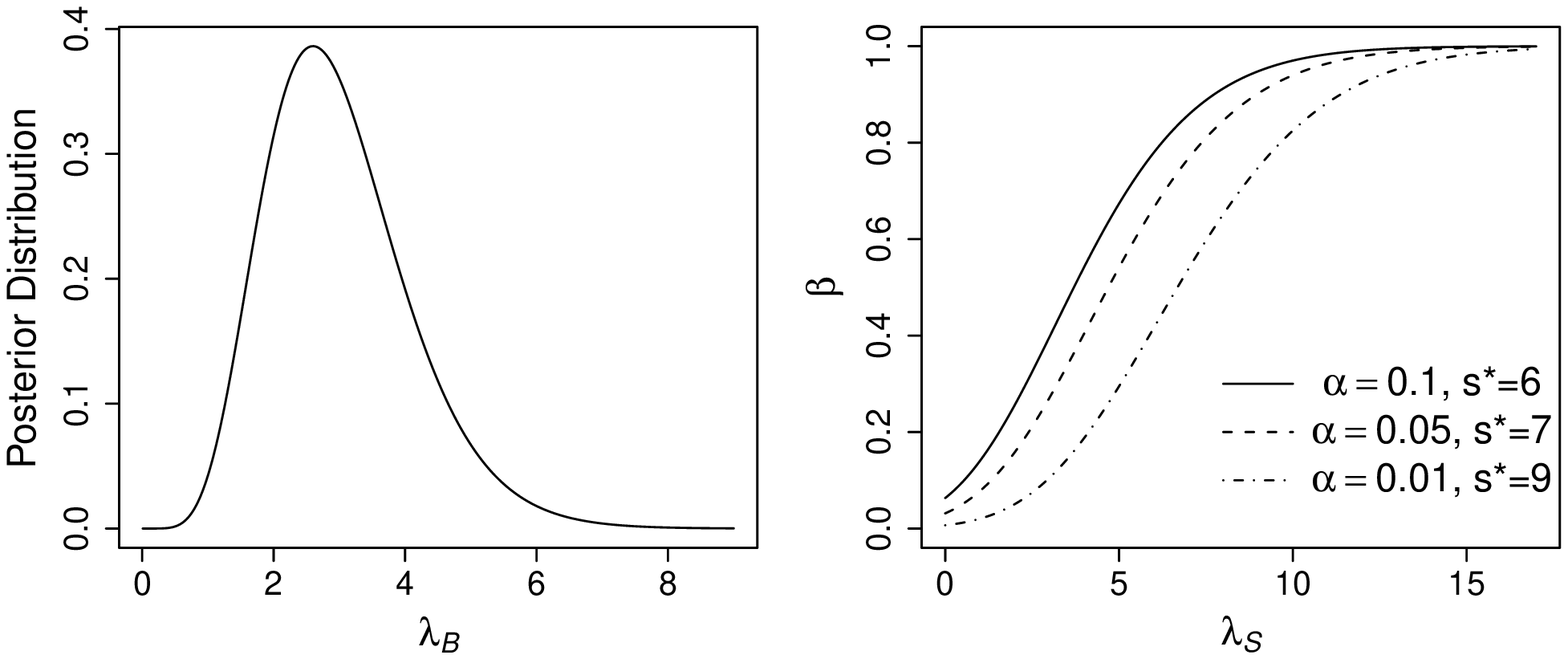} 
\caption{
Upper Limit with Unknown Background Intensity.  The left panel plots
a Bayesian posterior distribution for $\lamB$, $p(\lamB|\nB,\exptimeB,\rr)
\propto \lamB^{6.5} e^{-\lamB/2.5}$, that is used in
Equations~\ref{eq:ULpost} and \ref{eq:alpha2} to compute $\beta$,
the expected detection probability, as a function of $\lamS$ with
$\exptime=\exptimeB=\rr=1$.  The right panel plots $\beta(\lamS)$
for three values of $\alpha$ and their corresponding detection
thresholds.
}
\label{fig:unknown}
\end{center}
\end{figure*}

As an illustration, suppose the uncertainty in $\lamB$ given the
observed background counts can be summarized in the posterior
distribution plotted in the left panel of Figure~\ref{fig:unknown}.
This is a gamma posterior distribution of the sort that typically
arises when data are sampled from a Poisson distribution.  Using
this distribution, we can compute $\thresh$ for any given value of
$\alpha$ as the smallest value that satisfies Equation~\ref{eq:alpha2};
the results are given for three values of $\alpha$ in the legend
of Figure~\ref{fig:unknown}.  We then use Equation~\ref{eq:ULpost}
to compute $\beta(\lamS)$ as plotted in the right panel of
Figure~\ref{fig:unknown}.  The upper limit can be computed for any
$\beta_{\rm min}$ using these curves just as in Figure~\ref{fig:ulim}.

\subsection{Confidence Intervals versus Upper Limits}
\label{s:compare}

\begin{figure*}[p]
\begin{center}
\includegraphics[width=6.5in]{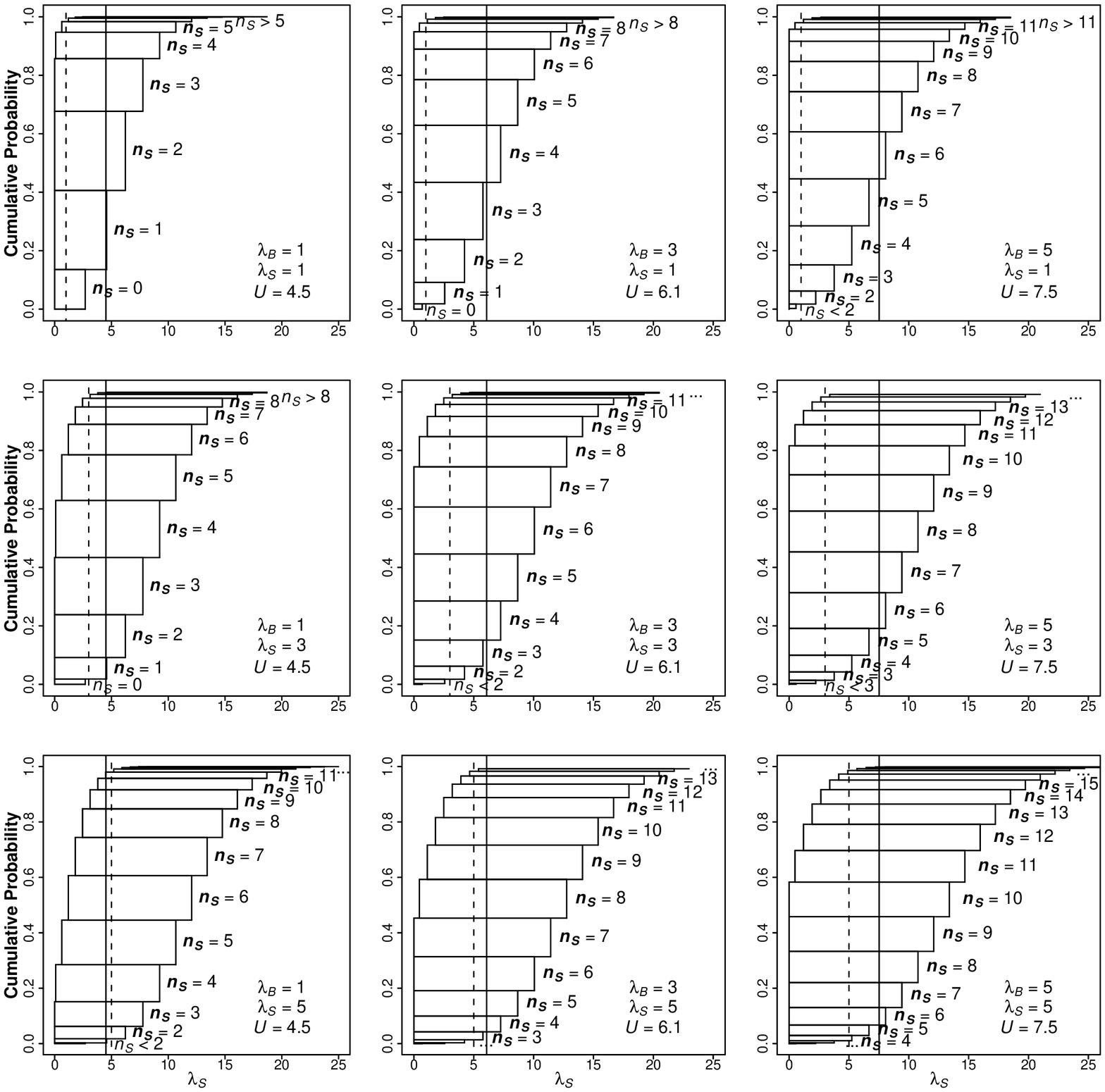} 
\caption{\small
Illustrating the Difference Between Confidence Intervals and Upper
Limits.  This figure is identical to Figure~\ref{fig:CI}, with an
additional solid vertical horizontal line showing the location of
the upper limit computed with $\beta_{\rm min}=0.8$.  The legend
denotes the true value of $\lamS$, the assumed known value of
$\lamB$, and the computed upper limit, $\ulim$.  Note that unlike
the confidence interval, which depends strongly on the number of
observed source counts, $\nS$, the upper limit is fixed once the
detection threshold $\thresh$ (which depends on $\lamB$) and the
minimum detection probability $\beta_{\rm min}$ are specified.
}
\label{fig:ulimCI}
\end{center}
\end{figure*}

Although the form of a confidence interval makes it tempting to use
its upper bound in place of an upper limit, this is misleading and
blurs the distinction between the power of the detection procedure
and the confidence with which the flux is measured.  As an illustration
we have computed the upper limit for each of the nine panels in
Figure~\ref{fig:CI} (using the true values of $\lamB$ reported in
each panel).  The results are plotted as solid vertical lines in
Figure~\ref{fig:ulimCI}.  Notice that unlike the upper bound of the
confidence interval the upper limit does not depend on $\nS$ and
takes on values that only depend on $\lamB$.  Using the upper bound
of the confidence interval sometimes overestimates and sometimes
underestimates the upper limit.  In all but one of the nine cases
with the highest $\lamS/\lamB$ (i.e., $\lamS=5$, $\lamB=1$) the
upper limit for $\lamS$ is larger than $\lamS$. Of the nine cases,
this is the one that is mostly likely to result is a source detection
and is the only one with a probability of detection greater than
$\beta_{\rm min}=0.8$.

\begin{figure*}[t]
\begin{center}
\includegraphics[width=6.5in]{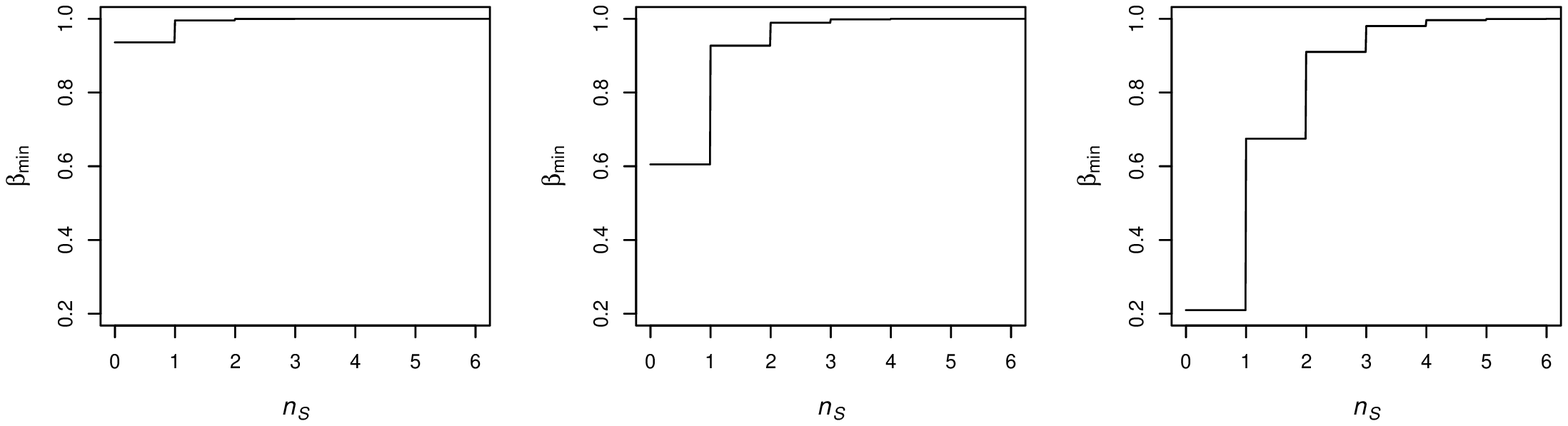} 
\caption{Interpreting Upper Bounds as Upper Limits.
The three panels plot the probability or source detection for a
source with $\lamS$ equal to the upper bound of Garwood's confidence
intervals.  Because the confidence intervals depend on $\nS$, the
detection probability, $\beta_{\rm min}$ increases with $\nS$.  The
three panels correspond to $\lamB=1$ (left), $\lamB=3$ (middle),
and $\lamB=5$ (right), as in the columns of Figure~\ref{fig:ulimCI}.
All calculations were preformed with $\exptime=\exptimeB=\rr=1$.
Because a source with intensity equal to the upper bound can have
a detection probability as low as 20\% or as high as 100\%,  the
upper bound does not calibrate the maximum intensity that a source
can have with appreciable probability of going undetected in any
meaningful way.
}
\label{fig:ub=ul}
\end{center}
\end{figure*}

Alternatively, we can compute the value of $\beta_{\rm min}$ required
for the upper bound of Garwood's confidence interval to be interpreted
as an upper limit.  Figure~\ref{fig:ub=ul} does this for the three
values of $\lamB$ used in the three columns of Figures~\ref{fig:CI}
and \ref{fig:ulimCI}.  Consider how the upper bound of Garwood's
confidence interval increases with $\nS$ in Figure~\ref{fig:ulimCI}.
Each of these upper bounds can be interpreted as an upper limit,
but with an increasing minimum probability of a source detection,
$\beta_{\rm min}$.  The three panels of Figure~\ref{fig:ulimCI}
plot how the required $\beta_{\rm min}$ increases with $\nS$ for
three values of $\lamB$.  Notice that a source with intensity equal
to the upper bound of Garwood's confidence interval can have a
detection probability as low as 20\% or essentially as high as
100\%.  Thus, the upper bound does not calibrate the maximum intensity
that a source can have with appreciable probability of going
undetected in any meaningful way.

\subsection{Statistical Selection Bias}
\label{s:eddington}

\begin{figure}[t]
\begin{center}
\includegraphics[width=2.9in]{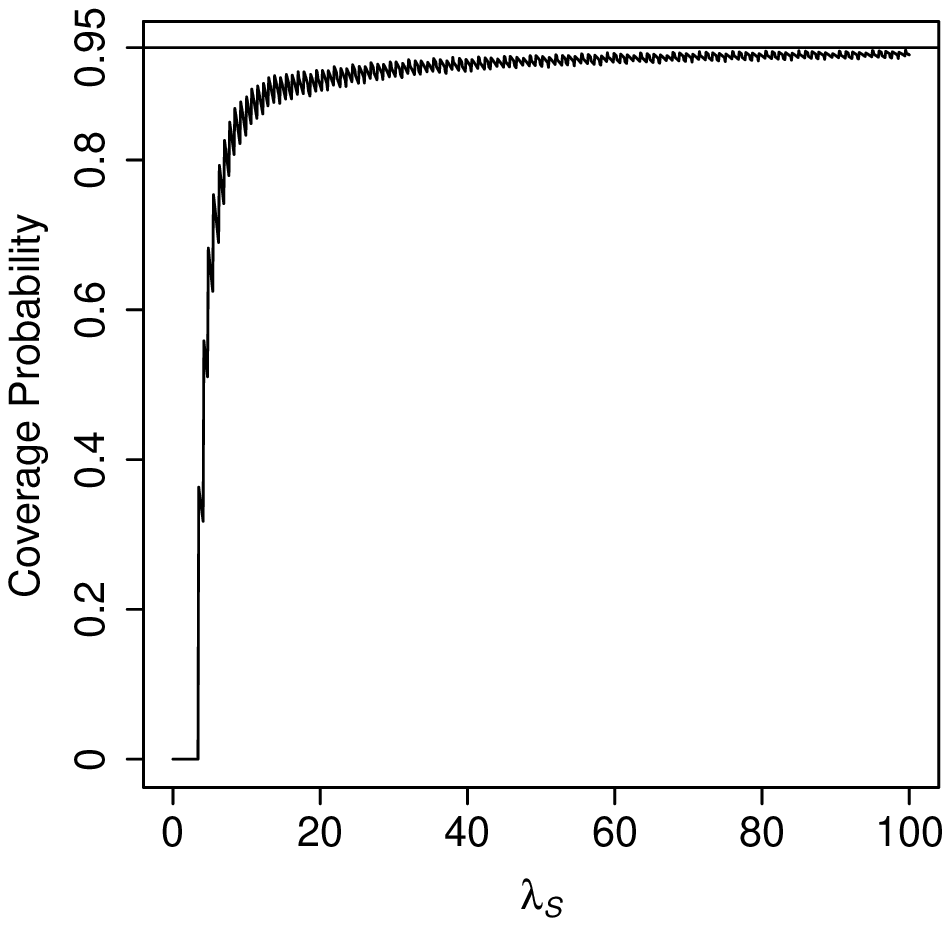} 
\caption{Conditional Coverage Probability of  Confidence Interval
Reported Only for Detected Sources.  When confidence intervals are
only reported for detected sources the coverage probability may be
very different than the nominal level of the interval.  This plots
shows the true coverage of the 95\% nominal intervals of Garwood
(1936) when they are only reported for sources detected with $\alpha
=0.05$ (with $\lamB=3$).  For small values of $\lamS$ the coverage
can be very low and as $\lamS$ grows, the coverage probability
converges to 95\%.
Far fewer intervals contain $\lamS$ than one would expect given the
nominal level.  Confidence intervals must be reported regardless
of $\nS$ and regardless of whether a source is detected.  The jagged
appearance of curve stems from the discrete nature of Poisson data.
}
\label{fig:cover}
\end{center}
\end{figure}

As mentioned in \S\ref{s:CI}.1 it is common practice to only report
a confidence interval for detected sources.  {\it Selectively
deciding when to report a confidence interval} in this way can
dramatically bias the coverage probability of the reported confidence
interval\footnote{A similar concern was raised by Feldman and Cousins
(1998) who noticed that deciding between a one-sided and a two-sided
confidence interval can bias the coverage probability of the resulting
interval, if the decision is based on the observed data.}.  We call
this bias a {\sl statistical selection bias}.
Note that this is similar to the Eddington bias (Eddington 1913)
that occurs when intensities are measured for sources close to the
detection threshold.  For sources whose intrinsic intensity is
exactly equal to the detection threshold, the average of the
intensities of the detections will be overestimated because downward
statistical fluctuations result in non-detections and thus no
intensity measurements.
In extreme cases this selection bias can lead to a nominal 95\%
confidence interval having a true coverage rate of well below 25\%,
meaning that only a small percentage of intervals computed in this
way actually contain $\lamS$.  As an illustration, Figure~\ref{fig:cover}
plots the actual coverage of the nominal 95\% intervals of Garwood
(1936) for a Poisson mean when the confidence intervals are only
reported if a  source is detected with $\alpha=0.05$.  These intervals
are derived under the assumption that they will be reported regardless
of the observed value of $\nS$.  Although alternative intervals
could in principle be derived to have proper coverage when only
reported for detected sources, judging from Figure~\ref{fig:cover}
such intervals would have to be wider than the intervals plotted
in Figure~\ref{fig:CI}.  {\it It is critical that if standard
confidence intervals are reported they be reported regardless of
the observed value of $\nS$ and regardless of whether a source is
detected.}\footnote{
Though this is usually not feasible when sources are detected via
an automated detection algorithm such as {\tt celldetect} or {\tt
wavdetect}.  However, in many cases, source detectability is
determined based on a pre-existing catalog, and in such cases, both
Limits and Bounds should be reported in order to not introduce
biases into later analyses.}

\subsection{The Detection Threshold as an Upper Limit}
\label{s:var}

As discussed in Section~\ref{s:intro}, the detection threshold is
sometimes used as an upper limit.  Under certain circumstances,
this can be justified under our definition of an upper limit.
Suppose that  some invertible function $f(\thrtest)$ can be used
as an estimate  of $\lamS$ and that for any $\lamB$, the sampling
distribution of $f(\thrtest)$ is continuous with median equal to
$\lamS$.
That is,
\begin{equation}
\prob\Big( f(\thrtest) > \lamS \Big| \lamS, \lamB, \exptime,  \exptimeB, \rr\Big) = 0.5.
\label{eq:DTasUL1}
\end{equation}
Because Equation~\ref{eq:DTasUL1} holds for any $\lamS$, it holds for $\lamS = f(\thresh)$.  That is,
\begin{equation}
\prob\Big( f(\thrtest) > f(\thresh) \Big| \lamS=f(\thresh), \lamB,  \exptime,  \exptimeB, \rr\Big) = 0.5.
\label{eq:DTasUL2}
\end{equation}
Inverting $f$ and integrating over $p(\lamB | \nB,\exptimeB,\rr)$, we have
\begin{equation}
\int \prob\Big( \thrtest > \thresh \Big| \lamS=f(\thresh), \lamB,  \exptime,  \exptimeB, \rr\Big) p(\lamB |\nB,\exptimeB,\rr)d\lamB = 0.5.
\label{eq:DTasUL3}
\end{equation}
Comparing Equation \ref{eq:DTasUL3} with Equation~\ref{eq:ULpost}
we see that $f(\thresh) = \ulim(\alpha, \beta_{\rm min} = 0.5)$.
Thus, if $f$ is an identity function the detection threshold  is
an upper limit.  Although the assumption that the sampling distribution
of $f(\thrtest)$ has median $\lamS$ for every $\lamB$, is unrealistic
in the Poisson case, it is quite reasonable  with Gaussian statistics.
Even if this assumption holds, $f(\thresh)$ is a weak upper limit
in that half the time a source with this intensity would go undetected
and there is a significant chance that sources with intrinsic
intensity larger than $f(\thresh)$  would remain undetected.

It should be emphasized that even when the detection threshold is
used as an upper limit, it is not an ``upper limit on the counts,''
but an upper limit on the intrinsic intensity of the source.  The
counts are an observed, not an unknown quantity.  There is no need
to compute upper bounds, error bars, or confidence intervals on
known quantities.  It is for the unknown source intensity that these
measures of uncertainty are useful.

\subsection{Recipe}
\label{s:recipe}

Our analysis of upper limits and confidence interval assumes that
the observables are photon counts that we model using the Poisson
distribution.  However, the machinery we have developed is applicable
to any process that uses a significance-based detection threshold.
Here, we briefly set out a general recipe to use in more complicated cases.  For complex detection algorithms, some of the steps may require Monte Carlo methods.
\begin{enumerate}
\item Define a probability model for the observable source and
background dataset given the intrinsic source  and background
strengths, $\lamS$ and $\lamB$, respectively.  For the simple Poisson
case, this is defined in Equation~\ref{eq:basicpoi}.  In many
applications, these could be approximated using Gaussian distributions.
It is typically required that a background data set be observed but
in some cases $\lamB$ may be known a priori.  The source could be
a spectral line, or an extra model component in a spectrum, or
possibly even more complex quantities that are not directly related
to the intensity of a source.
\item Define a test statistic $\thrtest$ for measuring the strength
of the possible source signal.  In the simple Poisson case, we set
$\thrtest=\nS$.  The ``source'' could be a spectral line or any
extra model component in a spectrum, or a more complex quantity
that is not related to the intensity of a source.
\item Set the maximum probability of a false detection, $\alpha$,
and compute the corresponding $\alpha$-level detection threshold,
$\thresh$.  Although $\thresh$ depends on $\lamB$, we can compute
the {\sl expected} $\thresh$ by marginalizing over $\lamB$ using
$p(\lamB|\nB)$ when $\lamB$ is not known exactly.  Likewise, if
$\lamS$ is defined as a function of several parameters, the same
marginalization procedure can be used to marginalize over any
nuisance parameters.  In this case, we typically marginalize over
$p(\eta | \nB)$ or perhaps $p(\eta |\nS,\nB)$, where $\eta$ is the
set of nuisance parameters.  In this regard, we are setting $\alpha$
to be a quantile of the posterior predictive distribution of
$\thrtest$, under the constraint that $\lamS=0$, see Gelman et
al.~(1996) and Protassov et al.~(2002).
\item Compute the probability of detection, $\beta(\lamS)$, for the
adopted detection threshold $\thresh$.
\item Define the minimum probability of detection at the upper
limit, $\beta_{\rm min}$.  Traditionally $\beta_{\rm min}=0.5$ has
been used in conjunction with $\alpha=0.003$ in astronomical analysis
(see \S\ref{s:var}).
\item Compute the smallest value of $\lamS$ such that $\beta(\lamS)\geq
\beta_{\rm min}$. This is the {\sl upper limit}.
\end{enumerate}

\section{Example: Signal to Noise Ratio}
\label{s:snr}

We focus below on Signal-to-Noise (SNR) based detection at a single
location as an example application.  The SNR was the primary statistic
used for detecting sources in high-energy astrophysics before the
introduction of maximum-likelihood and wavelet-based methods.
Typically, $\frac{S}{N}=3$ was used as the detection threshold,
corresponding to $\alpha=0.003$ in the Gaussian regime. Here we
apply the recipe in \S\ref{s:recipe} to derive an upper limit with
SNR-based detection.  Our methods can also be applied to more
sophisticated detection algorithms such as sliding-cell detection
methods such as {\tt celldetect} (Harnden et al.\ 1984, Dobrzycki
et al.\ 2000, Calderwood et al.\ 2001), and wavelet-based detection
methods such as {\tt pwdetect} (Damiani et al.\ 1997), {\tt zhdetect}
(Vikhlinin et al.\ 1997), {\tt wavdetect} (Freeman et al.\ 2002),
etc.  Implementation of our technique for these methods will vary
in detail, and we leave these developments for future work.

We begin with a Gaussian probability model for the source and background counts (Step 1 in \S\ref{s:recipe})
\begin{eqnarray}
\nB | (\lamB, \rr, \exptimeB) &\sim& {\cal N}\left(\mu=\rr\exptimeB\lamB, \ \sigma=\sqrt{\rr\exptimeB\lamB}\right) \cr
\nS | (\lamS, \lamB, \exptime) &\sim& {\cal N}\left(\mu=\exptime(\lamS +\lamB), \ \sigma=\sqrt{\exptime(\lamS+\lamB)}\right) \,,
\label{eq:snr-model}
\end{eqnarray}
where $\lamB$ and $\lamS$ are non-negative.  We assume that the source is entirely contained within the source cell and that the PSF does not overlap the background cell.  We can estimate $\lamB$ and $\lamS$ by setting $\nB$ and $\nS$ to their expectations (method of moments), as
\begin{equation}
\hat\lamB = {\nB\over r\exptimeB} \ \ {\rm and} \  \ \hat\lamS = {\nS\over\exptime} - {\nB\over r\exptimeB}.
\end{equation}
The variance of $\hat\lamS$ is
\begin{equation}
{\rm var}(\hat\lamS) = {\lamS\over \exptime} + {(\exptime+\rr\exptimeB)\lamB\over \rr\exptime\exptimeB}
\end{equation}
which we can estimate by plugging in $\hat\lamS$ and $\hat\lamB$ as
\begin{equation}
\widehat{\rm var}(\hat\lamS) = {\nS\over\exptime^2} + {\nB\over \rr^2\exptimeB^2}.
\end{equation}
To use the SNR as a detection criterion, we define (Step 2 in \S\ref{s:recipe})
\begin{equation}
\thrtest = {\hat\lamS\over \sqrt{\widehat{\rm var}(\hat\lamS)}} =  \frac{\rr\exptimeB\nS - \exptime\nB}{\sqrt{\rr^2\exptimeB^2\nS+\exptime^2\nB}}
\end{equation}
Step~3 in \S\ref{s:recipe} says that the maximum probability of a false detection should be set and $\thresh$ computed accordingly.  Instead we adopt the standard detection threshold, $\thresh =3$, used with the SNR and compute the corresponding $\alpha$,
\begin{equation}
\alpha(\lamB) = \int_{\cal R} p(\nB\rep | \lamB,\rr, \exptimeB) p(\nS\rep | \lamS =0, \lamB, \exptime) d\nB\rep d\nS\rep,
\label{eq:snr-alpha}
\end{equation}
where ${\cal R}$ is the region where $\thrtest(\nS\rep,\nB\rep)> \thresh = 3$ (see footnote~\ref{ft:poialpha}).  For given values of $\lamB$ and $\rr$, the integral in Equation~\ref{eq:snr-alpha} can be easily evaluated via Monte Carlo.  Alternatively, we can compute $\alpha$ by marginalizing over $\lamB$ if it is unknown,
\begin{equation}
\alpha = \int_0^\infty \alpha(\lamB) p(\lamB | \nB,\rr,\exptimeB) d\lamB,
\label{eq:snr-alpha2}
\end{equation}
where $\nB$ is the observed background count and $\alpha(\lamB)$
is computed in Equation~\ref{eq:snr-alpha}.  The probability of
detection, $\beta(\lamS$) is computed (Step 4 in \S\ref{s:recipe})
by evaluating the same integral as in Equation~\ref{eq:snr-alpha}
except that $\lamS$ is not set to zero.  We can make the same
substitution in Equation~\ref{eq:snr-alpha2} if $\lamB$ is unknown.
With $\beta_{\rm min}$ in hand (Step 5 in \S\ref{s:recipe}), we can
find the value of $\lamS$ such that $\beta(\lamS) =\beta_{\rm min}$
(Step 6 in \S\ref{s:recipe}).  This is the upper limit.

\begin{figure*}[p]
\begin{center}
\includegraphics[width=6.5in]{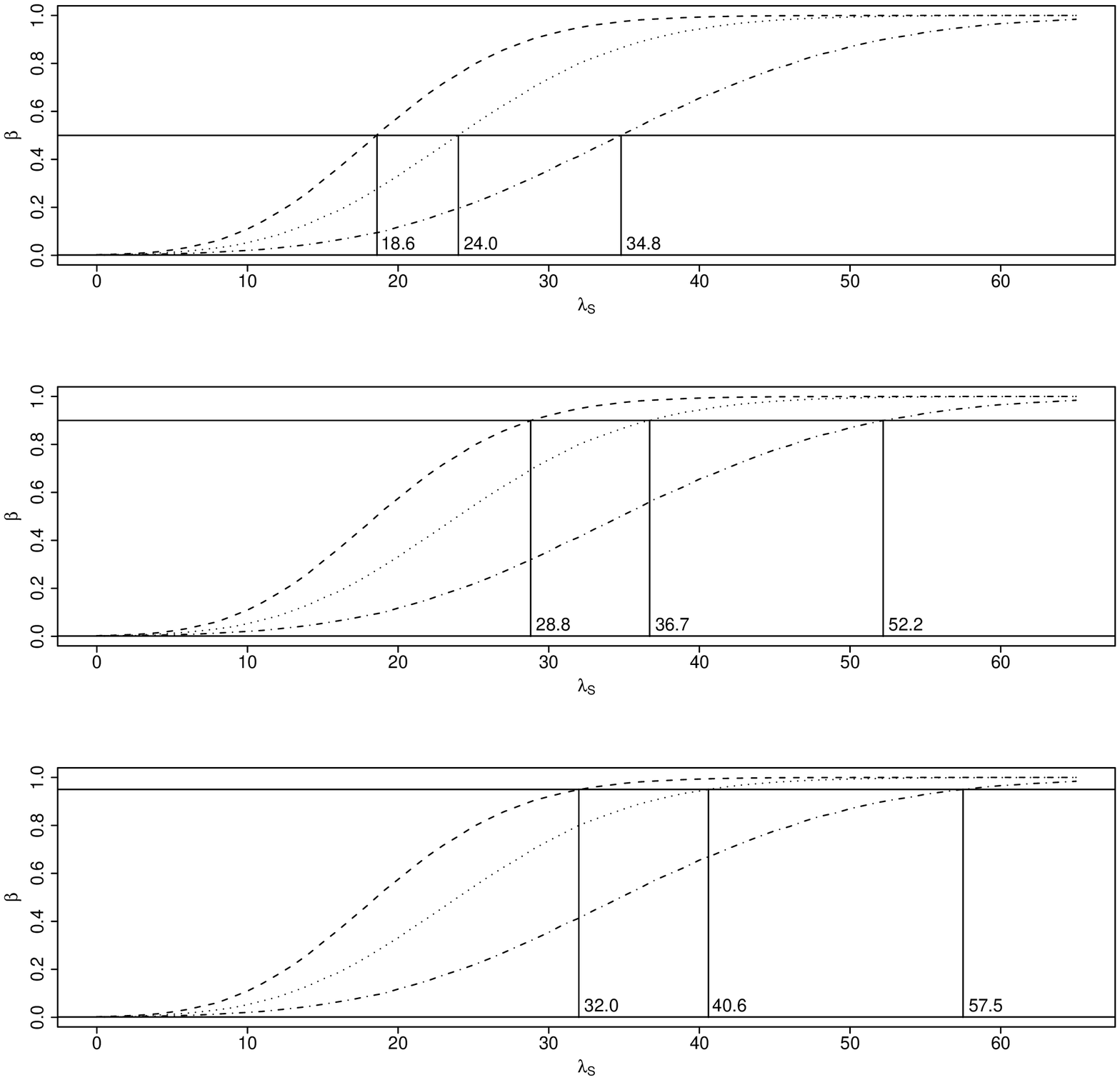} 
\caption{
Computing Upper Limits Based on the Probability of SNR Detection
of a Source.  The curves in each panel correspond to the probability
of source detection as a function of $\lamS$ using an SNR detection
threshold of $\thresh=3$.  The curves were computed with $\rr=1$,
$\exptime=1$, $\exptimeB=1$, and with $\lamB=10, 20,$ and $50$
(dashed, dotted, and dash-dotted lines, respectively).  Upper limits
are computed by first adopting an acceptable probability for a
source detection, and then computing the intercept on $\lamS$ of
the $\beta(\lamS)$ curves.  The panels show the value of the upper
limits for the different values of  $\lamB$ for $\beta_{\rm min}=0.5$
(top), $\beta_{\rm min}=0.9$ (middle), and $\beta_{\rm min}=0.95$
(bottom).
}
\label{fig:snr-power}
\end{center}
\end{figure*}

Figure~\ref{fig:snr-power} illustrates the use of
Equation~\ref{eq:snr-alpha} to compure $\beta(\lamB)$ for several
values of $\lamB$, with $\rr=\exptime=\exptimeB=1$.  The upper limit
is computed as the value of $\lamS$ such that $\beta(\lamS) =\beta_{\rm
min}$.  The three panels of Figure~\ref{fig:snr-power} report the
resulting upper limits for $\beta_{\rm min}= 0.5, 0.9,$ and $0.95$,
respectively.

\section{Summary}
\label{s:summary}

We have carefully considered the concept of upper limits in the
context of undetected sources, and have developed a rigorous formalism
to understand and express the concept. Despite its seeming simplicity,
upper limits are not treated in a uniform fashion in astronomical
literature, leading to considerable variations in meaning and value.
We formally define an upper limit to the source intensity as the
maximum intensity it can have without exceeding a specified detection
threshold at a given probability. This is defined by the statistical
power of the detection algorithm.  This is equivalent to defining
it as the largest source intensity that remains undetected at the
specified probability, and is defined by the probability of Type
II error.  Thus, if the detection probability is computed for a
variety of source intensities, the upper limit is then identified
by determining the intercept of the required probability with this
curve.  Thus, an upper limit is dependent only on the detection
criterion, which is generally a function only of the background,
and independent of the source counts. This is different from the
upper bound (i.e, the upper edge of a confidence interval), which
is obtained when the probability distribution of the source intensity
is computed given that some counts are observed in the putative
source region. We distinguish between the upper bound of the
confidence interval and the upper limit of source detectability.
Unlike a confidence interval (or Bayesian credible interval), an
upper limit is a function of the detection procedure alone and does
not necessarily depend on the observed source counts.

The primary goals of this paper are to clearly define an upper
limit, to sharpen the distinction between an upper limit and an
upper bound, and to lay out a detailed procedure to compute the
former for any detection process. In particular, we have shown how
to compute upper limits for the simple Poisson case.  We also provide
a step-by-step procedure for deriving it when a simplified
significance-based detection method is employed. To extract the
most science from catalogs, we argue for using a consistent,
statistically reasonable recipe of an upper limit being related to
the statistical power of a test. In addition, we illustrate the
peril of using an upper bound in place of an upper limit and of
only reporting a frequentist confidence interval when a source is
detected. Conversely, including confidence bounds, even for
non-detections, is a way to avoid the Eddington bias and increase
the scientific usefulness of large catalogs.

We also describe a general recipe for calculating an upper limit
for any well-defined detection algorithm.  Briefly, the detection
threshold should be first defined based on an acceptable probability
of a false detection (the $\alpha$-level threshold), and an intensity
that ensures that the source will be detected at a specifed probability
(the $\beta$-level detection probability) should be computed; this
latter intensity is identified with the upper limit.  We recommend
that when upper limits are reported in the literature, both the
corresponding $\alpha$ and $\beta$ values should also be reported.

\acknowledgments
This work was supported by NASA-AISRP grant NNG06GF17G (AC), CXC
NASA contract NAS8-39073 (VLK, AS), NSF grants DMS 04-06085 and DMS
09-07522 (DvD). We acknowledge useful discussions with Rick Harnden,
Frank Primini, Jeff Scargle, Tom Loredo, Tom Aldcroft, Paul Green,
Jeremy Drake, and participants and organizers of the SAMSI/SaFeDe
Program on Astrostatistcs.

\bigskip
\bigskip
\begin{center}
{\bf\Large Appendix}
\end{center}

\appendix

\section{Constructing a Confidence Interval by Inverting a Hypothesis Test}
\label{a:invert}

Here we discuss the relationship between confidence intervals and
hypothesis tests and in particular how a hypothesis test can be
used to construct a confidence interval.

A confidence interval reports the set of values of the parameter
that are consistent with the data.  When this set includes $\lamS=0$
it means that the data are consistent with no source and we expect
the null hypothesis not to be rejected and no source to be detected.
There is a more formal relationship between confidence intervals
and hypothesis testing and we can use a detection method to generate
a confidence interval.  Suppose that rather than testing the null
hypothesis that $\lamS=0$, we are interested in testing the more
general null hypothesis that $\lamS\leq\lamS^\star$, where $\lamS^\star$
is any non-negative number.  That is, we are interested in detecting
only sources of at least a certain brightness.  In this case, the
detection threshold, $\thresh(\lamS^\star)$, is defined as the
smallest value such that
\begin{equation}
\prob(\thrtest > \thresh(\lamS^\star) | \lamS=\lamS^\star,\lamB,\exptime,\exptimeB,\rr)\leq \alpha.
\label{eq:alphaint}
\end{equation}
Given an observed value of $\thrtest$, we can construct the set of
values $\lamS^\star$ for which we cannot reject the null hypothesis
that  $\lamS\leq\lamS^\star$.  This is a set of values of $\lamS$
that are consistent with the data and they form a $100(1-\alpha)$\%
confidence interval.  This particular, confidence interval, however
is of the form $(a, +\infty)$: For any observed count there is a
$\lamS^\star$ large enough so that we cannot reject the null
hypothesis that $\lamS\leq\lamS^\star$.  By reversing the null
hypothesis to $\lamS\geq\lamS^\star$ we can obtain an interval of
the form $(0, a)$ and by setting up a {\sl two-sided test} of the
null hypothesis that $\lamS=\lamS^\star$ against the alternative
hypothesis that $\lamS\neq\lamS^\star$ we can obtain an interval
of the more common form $(a, b)$.

\section{The Relationship between Upper Limits and the Power of the Test}
\label{a:ul-pow}

\begin{figure}[t]
\begin{center}
\includegraphics[width=2.9in]{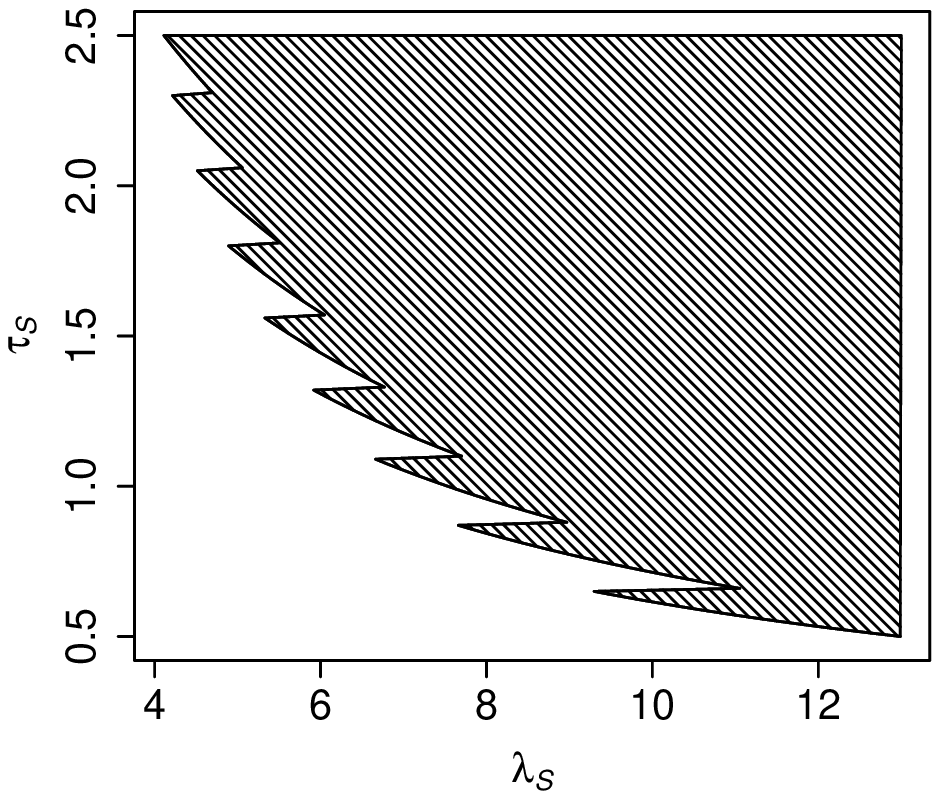} 
\caption{Dependence of the Upper Limit on the Exposure Time.
The shaded area to above and to the right of the curve is where the
detection probability exceeds $\beta_{\rm min} = 0.90$.  Thus the
curves give the upper limit (on the horizontal scale) as a function
of the exposure time.  The plot was made with $\lamB=3$, $\rr=5$,
$\exptime=\exptimeB$, and $\alpha=0.05$ and shows how the upper
limit generally decreases as the exposure time increases.  Because
of the discrete nature of Poisson data, the probability of type one
error can not be set exactly equal to $\alpha$.  This results in
the step function nature of $\alpha$ in Figure~\ref{fig:thresh} and
the non-monotonic decrease of the upper limit as a function of
exposure time here.
}
\label{fig:timelamS}
\end{center}
\end{figure}

An upper limit turns around the usual use of the power of a test.
Power is ordinarily used to determine the exposure time required
to be sure that a source with intensity ${\lamS}_{\rm  min}$ or
greater has at least probability $\beta_{\rm min}$ of being detected.
That is, the smallest $\exptime$ is found that satisfies
Equation~\ref{eq:power} for any $\lamS \geq {\lamS}_{\rm min}$ and
with ${\lamS}_{\rm min}$ fixed in advance.  Thus, power is used to
design an observation so that we have at least a certain probability
of detecting a source of given brightness.  With an upper limit on
the other hand $\exptime$ is fixed and Equation~\ref{eq:power} is
solved for $\lamS$.  This is illustrated in Figure~\ref{fig:timelamS}
which plots $\exptime$ versus  $\lamS$ with fixed $\lamB$, $\exptimeB$,
$\rr$, $\alpha$, and $\beta_{\rm min}$ and shows what values of
$\exptime$ and $\lamS$ satisfy Equation~\ref{eq:power} in the simple
Poisson case.  The shaded area above and to the right of the curve
is where the detection probability exceeds $\beta_{\rm min}= 0.90$.
Thus the curves give the upper limit (on the horizontal scale) as
a function of the exposure time.  The upper limit generally decreases
as the exposure time $\exptime$ increases, but not monotonically.
Due to the discrete nature of Poisson data, the threshold value
$\thresh$ changes in integer steps to allow for the inequality in
Equation~\ref{eq:alpha} to be satisfied.  This behavior may be
graphically illustrated by considering the sketch of the relevant
quantities in Figure~\ref{fig:alfabetillus}.  As $\exptime$ increases,
$\thresh$ increases in steps, causing the probability of false
detection to abruptly fall and then smoothly increase to $\alpha$.
As the expected background in the source region increases, the upper
curve shifts to the right, thereby increasing the shaded area that
lies above the threshold value.  However, when the area of the
shaded region in the upper plot becomes larger than the tolerable
probability of a Type~I error, $\thresh$ must be increased by one
to reduce that probability.  As $\exptime$ increases, the lower
curve remains stationary while $\thresh$ is unchanging.  At this
stage, the upper limit, $\ulim(\beta)$ is set as that value of
$\lamS$ which ensures that the Type~II error is $\beta$ (see
Equation~\ref{eq:power}), and thus slowly decreases as $\exptime$
increases.  When $\thresh$ increases as a step function, the lower
curve shifts to the right in order to maintain the same value of
$\beta$, and the upper limit abruptly increases.

\section{An Alternative Method for an Unknown Background Contamination Rate}
\label{a:bkgd}

In the body of the article, we suggested  conditioning on $\nB$ in
order to effectively estimate $\lamB$ when it is unknown.  A different
strategy conditions instead on the total count $\nS+\nB$ in order
to remove $\lamB$ from the model.  This method is based on the
simple probabilistic result that if $X$ and $Y$ are independent
Poisson variables with means $\lambda_X$ and $\lambda_Y$, respectively,
then given $X + Y$ , the variable $X$ follows a binomial distribution.
Applying this result to $\nS$ and $\nB$ with Poisson models given
in Equation\ref{eq:basicpoi}, we have
\begin{equation}
\nS \ | \ (\nS +\nB, \lamS, \lamB, \rr,\exptime,\exptimeB)
\sim {\rm Binomial} \left(\nS+\nB, {\exptime(\lamS+\lamB) \over \exptime \lamS +(\exptime +\rr\exptimeB)\lamB}\right),
\label{eq:binom1}
\end{equation}
a binomial distribution with $\nS+\nB$ independent counts each with
probability $\exptime(\lamS+\lamB) /(\exptime \lamS +(\exptime
+\rr\exptimeB)\lamB)$ of being a source count.  Reparameterizing
$(\lamS,\lamB)$ via $\xiS\lamB =\lamS+\lamB$, Equation~\ref{eq:binom1}
becomes
\begin{equation}
\nS \ | \ (\nS +\nB, \lamS, \lamB, \rr, \exptime, \exptimeB) \sim {\rm Binomial} \left(\nS+\nB, {\xiS \over \xiS + r\exptimeB/\exptime}\right),
\label{eq:binom2}
\end{equation}
which does not depend on the unknown background intensity.  Here
$\xi_S = (\lamS +\lamB)/\lamB$ which is equal to one if there is
no source and grows larger for brighter sources.  Because
Equation~\ref{eq:binom2} does not depend on $\lamB$ it can be used
for direct frequency based calculations even when $\lamB$ is unknown.
In particular, a detection threshold can be computed based on a
test of the null hypothesis that $\xiS=1$, which is equivalent to
$\lamS=0$.  This is done using Equation~\ref{eq:alpha} with
$\thrtest=\nB$ and using distribution given in Equation~\ref{eq:binom2}
with $\xiS=1$.  In particular, we find the smallest $\thresh$ such
that $\prob(\nS > \thresh | \nS+\nB,\xiS=1,\rr, \exptime, \exptimeB)\leq
\alpha$.  With the detection threshold in hand, we can compute an
upper limit for $\xiS$ using Equation~\ref{eq:power}.  The upper
limit is the smallest $\xiS$ such that $\prob( \nS > \thresh |
\nS+\nB, \xiS, \rr,\exptime, \exptimeB)\geq\beta_{\rm min}$.
Unfortunately, this upper limit cannot be directly transformed into
an upper limit for $\lamS$ without knowledge of $\lamB$ since $\lamS
= \lamB(\xi_S-1)$.

\end{document}